\documentclass{article}
\author{Claire Levaillant}
\title{A novel approach to multi-image quantum encryption/decryption using qudits}

\newcommand{\la}{\lambda}
\newcommand{\lL}{\lceil log_2\,L\rceil}
\newcommand{\lM}{\lceil log_2\,M\rceil}
\newcommand{\lT}{\lceil log_4\,T\rceil}
\newcommand{\lTe}{\lceil log_8\,T\rceil}
\newcommand{\lTd}{\lceil log_2\,T\rceil}
\newcommand{\lQ}{\lceil log_4\,Q\rceil}
\newcommand{\nts}{\negthickspace}

\newcommand{\lra}{\longrightarrow}

\newcommand{\rb}{\rbrace}
\newcommand{\lb}{\lbrace}
\newcommand{\R}{\mathbb{R}}
\newcommand{\N}{\mathbb{N}}
\newcommand{\al}{\alpha}
\newcommand{\be}{\beta}
\newcommand{\x}{\xi}
\newcommand{\ze}{\zeta}
\newcommand{\ga}{\gamma}
\newcommand{\io}{\iota}
\newcommand{\si}{\sigma}
\newcommand{\ta}{\tau}
\newcommand{\lf}{\lfloor}
\newcommand{\rf}{\rfloor}

\usepackage{amsmath, amssymb, amsthm, epsfig, graphicx}

\begin{document}
\maketitle

\noindent \textbf{Abstract. We introduce groundbreaking techniques in image encryption, assuming the existence of quantum computing ressources functioning with qudits, where $d$ is a power of $2$. Our quantum representation of color multi-image is based on space-filling curves and allows to reduce the storage space. We generalize the quantum baker map, so that it may scramble two $n$-qudits. By doing so, we enlarge its parameter space exponentially, leading to a better security. We define two new concepts of ``mixed scrambling" and ``mixed diffusion", and present a variety of schemes, depending on the needs of the users. }

\section{Introduction}

The past two decades has seen the field of quantum image encryption evolve rapidly. A quantum computer is used to encrypt one or multiple images after these have been put into a quantum representation. The ciphertext image or multi-image is put back into classical form using measurements and sent to the receiver who puts it back into quantum representation and applies the inverse quantum gates in the reverse order, using the private keys previously exchanged with the transmitter. By allowing massive parallel processing and increased data security, quantum computing has become the new approach for encrypting digital images. Because most quantum computers are built in order to allow for quantum computation with qubits, quantum representations for images have been developed using qubits. However, in recent years, several authors \cite{DON}\cite{MON} have started addressing quantum image encryption using qutrits instead of qubits. They consider images of size $3^n\times 3^n$ and each pixel position gets represented using $2n$ qutrits. They exhibit ternary quantum circuits that are used to prepare the images into quantum representation. An interest in using qutrits instead of qubits is based on bettering the hardware complexity. For instance, in the bit plane representation model of \cite{BA1}, eight bit planes are needed in order to encode a gray-scale intensity ranging from $0$ to $255$ and three qubits are used in order to encode the position of each bit plane. In contrast, in the qutrit representation model, for the same range of intensity, six trit planes are needed and thus, only two qutrits suffice in order to encode the positions of the trit planes. The same reasoning can be applied to show the storage efficiency of the pixel positions in the qutrit approach. In general quantum computing, most quantum compiling algorithms are designed using qubits instead of qutrits. Similarly, in quantum image encryption, most quantum transforms that are used for the scrambling of the images are implemented using qubits. Thus, the field of quantum image encryption using qutrits is at its early stage of advancement. 

The present paper is concerned with decreasing the storage space even more drastically. In order to do so, an idea will be to project each point of a plane, or a cube, or a hypercube to a single point of a line, following a space-filling curve which we will define. A space-filling curve is a continuous function whose domain is the interval $[0,1]$ and whose range reaches every point in a higher dimensional region such as the unit square or more generally an $n$-dimensional hypercube. Space-filling curves were originally discovered by Giuseppe Peano in the plane \cite{PEO}. The Peano curve passes through every point of the unit square. A year later, David Hilbert  had found its own curve and published it in \cite{HIL}. If by construction the maps are surjective, it is easy to show, see $\S\,2$, that any continuous map from the unit interval to the unit square cannot be bijective. Space-filling curves are part of the family of fractal curves discussed by Mandelbrot in \cite{MAN}. They are used in electronics \cite{BIA} and in combinatorics \cite{BAR} amongst other fields. In \cite{BAR}, the authors show heuristically how to reduce an optimization problem on the plane to one on the line. In \cite{BU1}\cite{BU2}, Butz explores a means of converging to a set of numbers in certain mathematical programming problems, where a conventional programming method is not possible. He shows that space-filling curves provide a tool for doing so. One advantage of a space-filling curve is that it provides an order in which some discrete points of the multi-dimensional space are being traversed by the curve. Space-filling curves can also be applied to multidimensional data mining or to finding solutions to N-body problems. Our space-filling curve will be inspired from the one of Schoenberg \cite{SCH} and achieves for our purpose what Schoenberg's curve cannot achieve. Schoenberg's construction is itself inspired from Lebesgue's space-filling curve. 

Given a $d$-dimensional hypercube of size $2^n$ where each integral coordinate is usually represented using $n$ qubits, and an integral point of the hypercube gets represented using $dn$ qubits, our construction allows to represent this integral point with an $n$-qutit where $t=2^d$. We will treat the case $d=2$ first, leading to a quantum representation for a multi-image with ququarts with decreased storage space. We then generalize the quantum baker map that is sometimes used for scrambling quantum images to ququarts. We will see that the latter map operates on a square of size $4^n\times 4^n$ and that the choice of admissible parameters is largely increased, compared to its original version which uses qubits. It results in a more secure scrambling phase that is less vulnerable to brute force attacks. Using the novel quantum representation and the generalized quantum baker map, we show how to perform an encryption/decryption of a color multi-image in the RGB color system. The diffusion is achieved using a 7D hyperchaotic system, cascaded with a sine chaotification together with chaos arising from Chebyshev transforms. A hyperchaotic system has several positive Lyapunov exponents, thus its behavior is harder to predict. The chosen discrete dynamical system has five positive Lyapunov exponents, thus has excellent pseudo-randomness performances. 
We next generalize our space-filling curve to any given dimension $d\geq 3$, that is we define a surjective and continuous map from the interval $[0,1]$ to the $d$-dimensional hypercube $[0,1]^d$ and explain its link to qutits, where $t=2^d$, in the framework of multi-image quantum encryption/decryption schemes. We further define a generalized quantum baker map on a square of size $t^n\times t^n$ which can shuffle two $n$-qutits bijectively. We then apply our results in the special case when $d=3$ and present several scrambling schemes for a quantum representation using quoctits. The techniques we invent allow for a better scrambling than in traditional quantum encryption/decryption schemes, including the most recent ones. We namely obtain a scrambling that is more efficient since various types of data can be all scrambled at once, and more powerful since the level of disorder is considerably increased and any correlation gets completely destroyed. 

Qutits have a theoretical existence in topological quantum computation. This mode of quantum computation is insensitive to errors caused by the environment because only the relative positions of the quasiparticles called ``anyons" matters, after these have been interchanged or ``braided" \cite{DAS}. We talk about fault-tolerant quantum computation \cite{KIT}. These quasiparticles are predicted to exist in fractional quantum Hall states. Quantum information is stored in states with multiple quasiparticles. For instance, a qutit with $t=2^d$ can be realized with four $SU(2)_{2^{d+1}-2}$ anyons of topological charge $2^d-1$. It can also be realized by four $SU(2)_{2^{d+1}-1}$ anyons of topological charges $2^d-1$, $2^d$, $2^d$, $2^d-1$. These are only two instances out of many options. Namely,  $SU(2)_k$ anyons obey the following fusion rules for a trivalent vertex with edges labelled by the topological charges $a,b,c$, modeling the fusion of two quasiparticles into a third quasiparticle \cite{ZWA}:
$$\left\lbrace\begin{array}{l}
a+b+c\;\text{is even}\\
a\leq b+c,\;b\leq a+c,\;c\leq a+b\\
a+b+c\leq 2k
\end{array}\right.$$

 Unitary operations are carried out by braiding the quasiparticles, fusing them and measuring them by Max-Zehnder or Fabry-Perot interferometry \cite{BOC}\cite{BON}.

While the physical realization of such quasiparticles is subject of active ongoing research for the low levels $k$ of $SU(2)_k$ Chern-Simons theory such as $k=2,3,4$ (Ising, Fibonacci, $SU(2)_4$ anyons), it is to expect that more progress will be made in the decades to come for higher levels as well. 

In order to perform any quantum computation on a quantum computer, we must be able to approach any unitary matrix up to desired accuracy. We then talk about ``universal quantum computation". A theorem by the couple Brylinski dating from $2001$ asserts that universal single qudit gates together with any $2$-qudit entangling gate, are sufficient for universal quantum computation. Building upon results of \cite{JON}, Freedman, Larsen and Wang have shown in \cite{FLW}\cite{FWE} that for a number $n\geq 3$ of quasiparticles and except when $k\in\lb 1,2,4\rb$ or $k=8$ and $n=4$, where $k$ is the level of the $SU(2)$ Chern-Simons theory, braiding alone suffices to approximate any special unitary of $SU(d)$, where $d$ is the dimension of the vector space spanned by a rooted fusion tree, where the root is the total charge of the quasiparticles, the $n$ leaves are indexed by the respective topological charges of the $n$ quasiparticles and the edges are indexed by the admissible fusion outcomes, following the fusion rules. 
Thus, for these values of $k$, single universality is achieved by braiding only. The respective levels $k=2^{d+1}-2$ and $k=2^{d+1}-1$ of interest here and mentioned in our discussion above, satisfy to this condition for single universality. However, a $2$-qudit entangling gate is still needed for universal quantum computation. The fact that such a gate would exist does not follow from the theorems of \cite{FLW}\cite{FWE}. Indeed, almost any entangling braid between the two qudits will introduce a nontrivial charge line in between them, which results in leaving the computational vector space formed by the two qudits. We say that it introduces ``leakage" between the two qudits, as these will continue leaking while the quantum computation continues unfolding. Indeed, some of the following braids performed during the quantum computation will continue augmenting the computational space, so that this leakage cannot be minimized, nor controlled. 
There is for instance active research on the existence or the non existence of leakage-free entangling braids with Fibonacci anyons \cite{CWA}. In \cite{BIG}, we made an exact leakage-free $2$-qubit entangling gate with Fibonacci anyons, where braiding is supplemented with measurement operations. Entangling gates on two qutrits have also been achieved in \cite{BSM} and independently \cite{CUI} at level $4$ in the Kauffman-Jones version of $SU(2)$ Chern-Simons theory \cite{KAU}. It is a reasonable expectation that such entangling gates exist at higher levels than $4$, which can be realized by braiding and measuring the anyons. Thus, there is hope for one day implementing color multi-image quantum encryption/decryption schemes with qudits, using anyonic systems. It is likely that other quantum computing approaches will also flourish using qudits with $d\geq 3$, instead of qubits. 

\section{Quantum encryption/decryption scheme with ququarts}
\subsection{Preliminaries}
We briefly recall below Schoenberg's construction on which our own construction will be based. Let $f(t)$ be the even function of period two defined over the interval $(0,1)$ by 
$$\left\lbrace\begin{array}{l}f(t)=0\;\text{over}\;(0,\frac{1}{3})\\
f(t)=1\;\text{over}\;(\frac{2}{3},1)\\
f(t)\;\text{is linear over}\;(\frac{1}{3},\frac{2}{3})\end{array}\right.$$
\begin{center}\epsfig{file=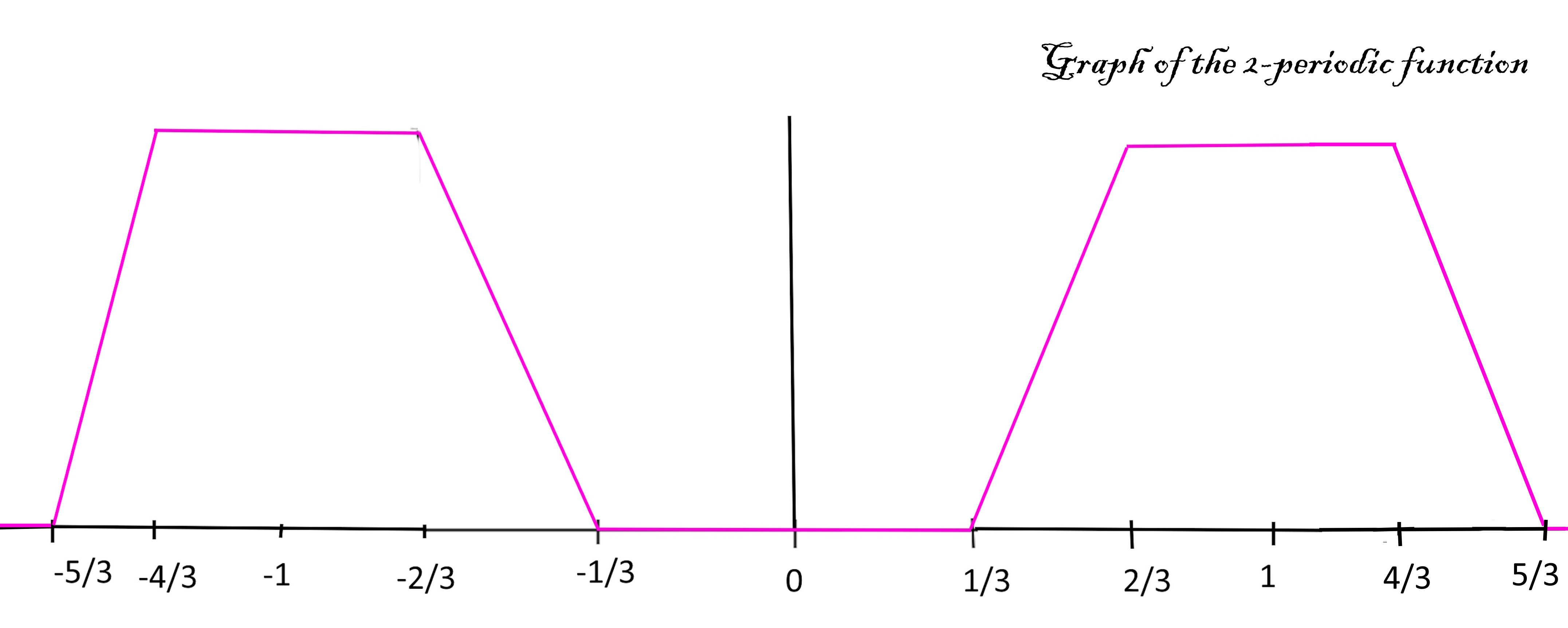, height=3.5cm}\end{center}
Schoenberg's curve is then defined over $[0,1]$ by:
$$x(t)=\sum_{k=1}^{+\infty}\frac{f(3^{2k-2}t)}{2^k},\;
y(t)=\sum_{k=1}^{+\infty}\frac{f(3^{2k-1}t)}{2^k}.$$
The two defining series of functions above both converge uniformly over $[0,1]$ as $f$ takes values betwen $0$ and $1$, implying the continuity of the limit functions which they define since $f$ is itself continuous. It remains to show the surjectivity of the map:
$$S:\begin{array}{ccc}[0,1]&\lra&[0,1]^2\\&&\\
t&\longmapsto& (x(t),y(t))\end{array}.$$
Let $(x,y)\in [0,1]^2$. Write:
$$x_0=\sum_{k=1}^{+\infty}\frac{a_{2k-2}}{2^k},\;y_0=\sum_{i=1}^{+\infty}\frac{a_{2k-1}}{2^k},$$
where for each $i\leq0,\;a_i\in\lbrace 0,1\rbrace$.
Define: $$t_0=\sum_{k=1}^{+\infty}\frac{2a_{k-1}}{3^k}.$$
Notice, if $a_0=0$ (resp $a_0=1$), then $0\leq t_0\leq \frac{1}{3}$ (resp $\frac{2}{3}\leq t_0\leq 1$) and so $f(t_0)=0$ (resp $f(t_0)=1$). Anyhow, we have $f(t_0)=a_0$. Moreover, we have for all integers $k\geq 1$:
$$3^k\,t_0=\text{even integer}\;+\sum_{i=1}^{+\infty}\frac{2a_{k+i-1}}{3^i},$$
from which we derive that $$f(3^k\,t_0)=a_k,$$ by the previous reasoning. It comes $x(t_0)=x_0$ and $y(t_0)=y_0$. 

The fact that the map $S$ is not injective can be seen using for instance the fact that any rational number $\frac{a}{2^m}$ with $a$ odd integer such that $0<a<2^m$ has two distinct decompositions in base $2$, one of which is stationary at $1$ and is usually called the improper decomposition. This fact is actually much more general. 
\newtheorem{Proposition}{Proposition}
\begin{Proposition}
There does not exist any continuous and bijective map $F:[0,1]\lra [0,1]^2$. 
\end{Proposition}
\textsc{Proof.} Suppose such a map $F$ exists. Since any closed set of a compact set is compact, $[0,1]$ is compact and $F$ is continuous, the direct image of a closed set under $F$ is a compact of $[0,1]^2$, in particular is closed. Then the reciprocal image of any closed set under $F^{-1}$ is closed, which implies that $F^{-1}$ is continuous. Let $c\in (0,1)$. The set $[0,1]^2-\lb F(c)\rb$ is path-connected, hence is connected. However, 
\begin{eqnarray}F^{-1}([0,1]^2-\lb F(c)\rb&=&F^{-1}([0,1]^2)-\lb c\rb\\
&=&[0,1]-\lb c\rb\\
&=&[0,c)\cup (c,1),\end{eqnarray}
and the set to the right hand side of Eq. (3) is not connected, thus providing a contradiction. \hfill $\square$

In the framework of quantum image processing, we note however that Schoenberg's construction does not allow to reduce the number of qubits needed. Indeed, suppose we have an image of size $2^n\times 2^n$. Pixel coordinates are couples $(x,y)$ of integers belonging to $\lb 0, 1,\dots,2^n-1\rb$, and so $\big(\frac{x}{2^n},\frac{y}{2^n}\big)\in [0,1]^2$. 
If $\overline{a_{n-1}\dots a_1a_0}$ and $\overline{b_{n-1}\dots b_1b_0}$ are the respective decompositions of $x$ and $y$ in base $2$, we have:
\begin{equation}\frac{x}{2^n}=\sum_{i=0}^{n-1}\frac{a_i}{2^{n-i}}, \;\frac{y}{2^n}=\sum_{i=0}^{n-1}\frac{b_i}{2^{n-i}}, \end{equation}
with all the $a_i$'s and $b_i$'s taking their values in $\lb 0,1\rb$. 
The pixel $(x,y)$ can be represented using $2n$ qubits. By the above, 
$$t=\sum_{k=1}^n\Big(\frac{2a_{n-k}}{3^{2k-1}}+\frac{2b_{n-k}}{3^{2k}}\Big)$$ is a pre-image of $(x,y)$ under $S$. We see that $2n$ qubits are still needed in order to represent $t$. Thus, we have decreased the space dimension without bettering the storage. 
A remedy consists of introducing two distinct functions in the defining series instead of simply one. This is the purpose of the next section. 

\subsection{The space-filling curve}
Our space-filling curve is defined in the following theorem.  
\newtheorem{Theorem}{Theorem}
\begin{Theorem} Let $f,g:\R\lra\R$ be two continuous $1$-periodic functions such that:
$$\left\lbrace\begin{array}{l}
\forall t\in[0,\frac{1}{8}],\;f(t)=g(t)=0\\\\
\forall t\in[\frac{2}{8},\frac{3}{8}],\;f(t)=0,\,g(t)=1\\\\
\forall t\in[\frac{4}{8},\frac{5}{8}],\;f(t)=1,\,g(t)=0\\\\
\forall t\in[\frac{6}{8},\frac{7}{8}],\;f(t)=1,\,g(t)=1
\end{array}.\right.$$
The two functions $f$ and $g$ can for instance be linear elswehere. \\
\begin{center}\epsfig{file=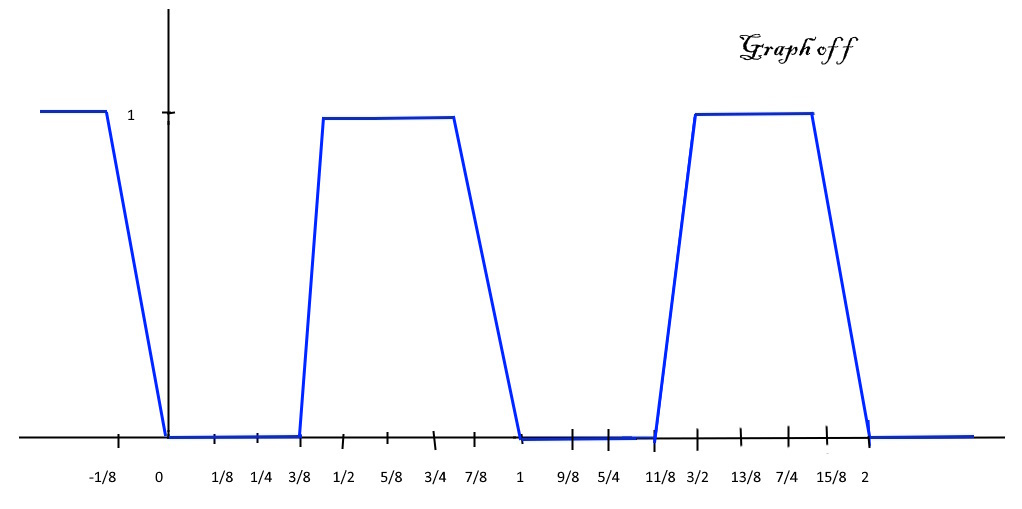, height=5.3cm}\end{center}
\begin{center}\epsfig{file=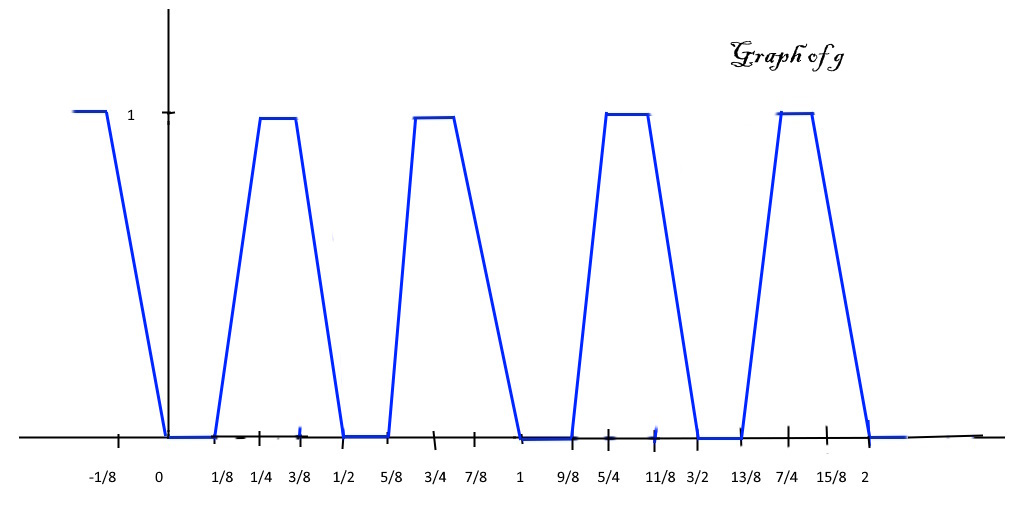, height=5.3cm}\end{center}
Define $$u(t)=\sum_{n=1}^{+\infty}\frac{f(8^{n-1}t)}{2^n},\; v(t)=\sum_{n=1}^{+\infty}\frac{g(8^{n-1}t)}{2^n}.$$
Then the map $$T:\begin{array}{ccc}[0,1]&\lra& [0,1]^2\\
t&\longmapsto &(u(t),v(t))\end{array}$$
is a space-filling curve. 
\end{Theorem}
\textsc{Proof.} The maps $u$ and $v$ are both continuous by uniform convergence of the two series defining them and the fact that $f$ and $g$ are both continuous. Thus, $T$ is continuous. It remains to show that $T$ is surjective. 

Let $t\in [0,1]$ and decompose
$$t=\sum_{n=1}^{+\infty}\frac{t_n}{8^n},$$
where the $t_n$'s are octits. 
We have:
$$\forall n\geq 1,\;8^{n-1}t=N_n+\frac{t_n}{8}+R_n,$$
with $$N_n\in\N\;\text{and}\;R_n=\sum_{i=n+1}^{+\infty}\frac{t_i}{8^{i+1-n}}\in \Big[0,\frac{1}{8}\Big].$$
Then, since $f$ and $g$ are $1$-periodic, we get:
$$\forall n\geq 1,\;\left\lb\begin{array}{cccc}2^{-n}f(8^{n-1}t)&=&2^{-n}f\big(\frac{t_n}{8}+R_n\big),&\text{with $R_n\in \big[0,\frac{1}{8}\big]$}\\&&&\\2^{-n}g(8^{n-1}t)&=&2^{-n}g\big(\frac{t_n}{8}+R_n\big),&\text{with $R_n\in \big[0,\frac{1}{8}\big]$}
\end{array}\right.$$
From there, given any $(u,v)\in [0,1]^2$, decompose uniquely $u$ and $v$ as 
$$u=\sum_{n=1}^{+\infty}\frac{u_n}{2^n},\;v=\sum_{n=1}^{+\infty}\frac{v_n}{2^n},$$
where $u_n$ and $v_n$ are all bits and using the proper expansion in the case when $u$ or $v$ admit exactly two such decompositions. Then, we construct a pre-image of $(u,v)$ under $T$ as follows:

$$\begin{array}{l}\text{- If $(u_n,v_n)=(0,0)$, set $t_n:=0$;}\\
\text{- If $(u_n,v_n)=(0,1)$, set $t_n:=2$;}\\
\text{- If $(u_n,v_n)=(1,0)$, set $t_n:=4$;}\\
\text{- If $(u_n,v_n)=(1,1)$, set $t_n:=6$.}\end{array}$$

Given an image of size $2^n\times 2^n$ and a pixel position $(x,y)$, decompose 
\begin{equation}
\frac{x}{2^n}=\sum_{k=1}^{n}\frac{a_{n-k}}{2^k}, \;\frac{y}{2^n}=\sum_{k=1}^{n}\frac{b_{n-k}}{2^k}, 
\end{equation}
using the same notations as in Eq. (4), and where the bits $a_i$'s and $b_i$'s are uniquely determined by the respective binary decompositions of the integers $x$ and $y$. Then, for each $k\in\lb 1,\dots, n\rb$, define 
$t_{n-k}:=0$ (resp $t_{n-k}:=2$, resp $t_{n-k}:=4$, resp $t_{n-k}:=6$) if $(a_{n-k},b_{n-k})=(0,0)$ (resp $((a_{n-k},b_{n-k})=(0,1)$, resp $(a_{n-k},b_{n-k})=(1,0)$, resp $(a_{n-k},b_{n-k})=(1,1)$). Then, $t_0$, $t_1$, $\dots$, $t_{n-1}$ are $n$ quarts representing the pixel position $(x,y)$. Thus, a pixel position can be represented using $n$ ququarts $|0>$ (when $t_i=0$), $|1>$ (when $t_i=2$), $|2>$ (when $t_i=4$), $|3>$ (when $t_i=6$). 
\subsection{The QQRMI and the QQQPCMIR}
The quantum quarts representation for a multi-image (abbreviated QQRMI) and the quantum quarts quart plane color multi-image representation (abbreviated QQQPCMIR) allow to decrease the storage needs while allowing an efficient scrambling of the pixel positions and of a large set of images, all at once. 

Given $4^n$ images of size $2^n\times 2^n$, we have seen that the pixel positions can be represented using $n$ ququarts. The images numbered $0,1,\dots, 4^n-1$ can themselves be represented using $n$ ququarts. Then, the QQRMI of the multi-image M reads:
\begin{equation}
|M>=\frac{1}{4^n}\sum_{m,z=0}^{4^n-1}|mz>|P_{mz}^{(4)}P_{mz}^{(3)}P_{mz}^{(2)}P_{mz}^{(1)}>.
\end{equation}
In this representation, the usual $8$ bit planes allowing an intensity ranging from $0$ to $255$ are divided into four quart planes: we group the bit planes two by two, for instance the two bit planes corresponding to the two least significant bits are paired together and the two bit planes corresponding to the two most significant bits are paired together. In between, the remaining four bit planes get also grouped by pairs. The quart value on a quart plane is $|0>$ (resp $|1>$, resp $|2>$, resp $|3>$) if the two bit values on the associated bit planes are $(0,0)$ (resp $(0,1)$, resp $(1,0)$, resp $(1,1)$), where the first bit in these couples refers to the most significant bit. In case the number of images to encrypt is even larger or much larger than $4^n$, we form $T$ blocks of $4^n$ images, with the last block possibly filled with blank images. Then, the quantum block quarts representation of multi-image, abbreviated QBQRMI, reads instead:
\begin{equation}
|M>=\frac{1}{2^{2n+\lT}}\sum_{m,z=0}^{4^n-1}\sum_{t=0}^{4^{\lT}-1}|t>|mz>|P_{mz}^{(4)}P_{mz}^{(3)}P_{mz}^{(2)}P_{mz}^{(1)}>.
\end{equation}
In the representation above, $m$ is the position of the image inside the $t$-th block. 
The two representations above are similar to and generalize the NEQR representation of \cite{NEQ} in that all the quart values on each quart plane get displayed. However, if we deal with color images instead of gray-scale images, then in order to limit the number of ququarts representing the quart values in the RGB system, we will use instead one ququart to represent the position of the four quart planes and three ququarts to represent the respective quart values of red, blue and green. This is the purpose of the QQQPCMIR representation which we define below as:
\begin{equation}
|M>=\frac{1}{4^n}\sum_{m,z=0}^{4^n-1}|mz>|q>|R>|G>|B>,
\end{equation}
where $|R>$, $|G>$ and $|B>$ are ququarts which can take the values $|0>$, $|1>$, $|2>$, $|3>$, depending on the quart value on the $q$-th quart plane. The $0$-th plane carries the least significant quart. 

Again, it is possible to form blocks of color images and the quantum blocks quarts quart plane color multi-image representation, abbreviated QBQQPCMIR, reads the following:
\begin{equation}
|M>=\frac{1}{2^{2n+\lT}}\sum_{m,z=0}^{4^n-1}\sum_{t=0}^{4^{\lT}-1}|t>|mz>|q>|R>|G>|B>. 
\end{equation}

Finally, all these representations generalize to any even number $2Q$ of bit planes, instead of the traditional number of eight bit planes. In that case, the respective four representations described above rewrite as follows.
\begin{eqnarray}
|M>&=&\frac{1}{4^n}\sum_{m,z=0}^{4^n-1}|mz>|P_{mz}^{(Q)}>\dots |P_{mz}^{(1)}>\\
|M>&=&\frac{1}{2^{2n+\lT}}\sum_{m,z=0}^{4^n-1}\sum_{t=0}^{4^{\lT}-1}|t>|mz>|P_{mz}^{(Q)}>\dots |P_{mz}^{(1)}>\\
|M>&=&\frac{1}{2^{2n+\lQ}}\sum_{m,z=0}^{4^n-1}\sum_{q=0}^{4^{\lQ}-1}|mz>|q>|RGB>\\
|M>&=&\lambda\nts\sum_{m,z=0}^{4^n-1}\nts\sum_{t=0}^{4^{\lT}-1}\sum_{q=0}^{4^{\lQ}-1}\nts\nts\nts\nts |t>|mz>|q>|RGB>,
\end{eqnarray}
where $$\lambda=\frac{1}{2^{2n+\lT+\lQ}}.$$
We note that both representations in $(12)$ and $(13)$ also apply to gray-scale images, where it suffices to replace the $3$-ququart $|RGB>$ with a single ququart $|P>$. 
While the QQRMI is similar to the novel enhanced quantum representation NEQR \cite{NEQ}, the plane approach of the QQQPCMIR is inspired from the bit plane representation for a quantum image, abbreviated BRQI, introduced by the authors of \cite{BA1}. In \cite{LE0}, we generalized the BRQI to multi-images and named the corresponding representation the bit plane representation for quantum multiple images, abbreviated BRQMI. In \cite{LE1}, we introduce a version by blocks, namely the quantum block bit plane representation for a multi-image, abbreviated by QBBRMI, so that the number of images in each block equals the number of bit planes, thus allowing a block independent scrambling of the bit planes and of the images belonging to a block. Because the quantum transform we use, namely the quantum baker map, operates on a square, the interest of using blocks was to avoid having to add many bit planes in the case when we need to encrypt a large number of images, well exceeding the number of bit planes. Instead, the quantum baker map gets applied on a square of size $2^{\lL}$, where $L$ denotes the number of bit planes. It allowed to save a substancial number of qubits. We will now draw a comparison of the bit models with their analog quart models. When the number of $2^n\times 2^n$ images exceeds the number of bit planes, the BRQMI uses $2n+2\lM+1$ qubits, where $M$ is the number of images to encrypt, see \cite{LE0}. Under the same assumptions, the QBBRMI reduces the number of qubits to $2n+\lM+\lL+1$, where $L$ is the (even) number of bit planes, see \cite{LE1}. Assuming now the number of images is $4^n$, these two numbers are respectively $6n+1$ and $4n+\lL+1$. Under the same notations, and still considering like in \cite{LE0} and \cite{LE1} gray-scale images, the ququart representation with quart planes, uses $2n+\lceil log_4\frac{L}{2}\rceil+1$ ququarts. Let us get even more specific in our comparison, namely suppose we have the traditional number of eight bit planes $(L=8)$ and the $2024$ images have  size $256\times 256$, then by using the most space efficient representation by blocks in the qubit representation, the latter representation uses a total of $36$ qubits; the ququarts representation with quart planes uses a total of $18$ ququarts (this number becomes $20$ if we rather use the NEQR type representation presented above as QQRMI). Suppose now that we have $8096$ images, all of size $512\times 512$. The qubit representation uses $40$ qubits, versus $20$ ququarts (or $22$ in NEQR type) for the ququart representation.

The choice of $4^n$ images or blocks of $4^n$ images is not random. Namely, having $n$ ququarts representing the images position and $n$ ququarts representing the pixel position will allow to scramble all the images and pixel positions using a generalized quantum baker map which we define in the next section. 
\subsection{A generalized quantum baker map on ququarts}
The quantum baker map is used for quantum scrambling because it has a longer period than other quantum transforms such as the quantum Arnold transform. Also, it has an easier quantum implementation using SWAP and controlled SWAP gates \cite{BA1}\cite{LE1}. The quantum baker map was introduced by the authors of \cite{BA0} to operate for a discretized square of size $2^n\times 2^n$ on two $n$-qubits by shuffling them. We presently generalize it, so that it now operates on a discretized square of size $4^n\times 4^n$ by scrambling two $n$-ququarts. 

\newtheorem{Prop}{Proposition}
\newtheorem{Definition}{Definition}

\begin{Definition}
The generalized discrete baker map with parameters $(q_1,\dots, q_k)$ is defined for a square of size $4^n\times 4^n$ with
$$4^n=4^{q_1}+\dots +4^{q_k}$$
as
$$\left\lbrace\begin{array}{l}
N_0=0\; \text{and}\; N_i=4^{q_1}+\dots +4^{q_i}\;\text{and}\;
x=N_{i-1},N_{i-1}+1,\dots,N_i-1\\
(x^{'},y^{'})=\Bigg(4^{n-q_i}(x-N_{i-1})+y\,\text{mod}\,4^{n-q_i},N_{i-1}+\frac{y-y\,\text{mod}\,4^{n-q_i}}{4^{n-q_i}}\Bigg)
\end{array}\right.$$
\end{Definition}

\begin{Prop}
(i) The map defined by
$$G_s(x,y)=\Bigg( 4^{n-s}x\,\text{mod}\,4^n+y\,\text{mod}\,4^{n-s},\frac{y-y\,\text{mod}\,4^{n-s}}{4^{n-s}}+x-x\,\text{mod}\,4^s\Bigg)$$
has a quantum implementation. \\
(ii) Each subfunction on $\lbrace N_{i-1},\dots,N_i-1\rbrace$ of the generalized discrete baker map is $G_{q_i}$ if and only if $4^{q_i}|4^{q_1}+\dots+4^{q_{i-1}}$.
\end{Prop}
\textsc{Proof.} (i) Write $x$ and $y$ in base $4$:
\begin{eqnarray*}
x&=&x_{n-1}\dots x_0\\
y&=&y_{n-1}\dots y_0
\end{eqnarray*}
Then, we have:
\begin{equation}
G_s(x,y)=(x_{s-1}\dots x_1x_0y_{n-s-1}\dots y_1y_0,x_{n-1}\dots x_sy_{n-1}\dots y_{n-s})
\end{equation}
Then, $G_s$ can be implemented using SWAP gates. 

(ii) Suppose that for all integers $x$ with $N_{i-1}\leq x\leq N_i-1$, the generalized baker map is $G_{q_i}$. It comes:
$$N_{i-1}=x-x\;\text{mod}\, 4^{q_i}$$
In particular, $$4^{q_i}|N_{i-1}=4^{q_1}+\dots 4^{q_{i-1}}.$$
Conversely, suppose this divisibility condition holds. Notice that since $N_{i-1}\leq x<N_i$, we have $0\leq x-N_{i-1}<4^{q_i}$, so that $0\leq 4^{n-q_i}x<4^n$. Then, from 
\begin{eqnarray*}
4^{n-q_i}(x-N_{i-1})&=&4^{n-q_i}x-\frac{N_{i-1}}{4^{q_i}}4^n,
\end{eqnarray*}
we derive:
$$4^{n-q_i}(x-N_{i-1})=(4^{n-q_i}x)\;\text{mod}\,4^n.$$
\hfill $\square$

It is straightforward to adapt the discussion of $\S\,3.1$ of \cite{LE1}, thus leading to a quantum circuit for the generalized quantum baker map, with parameters $(q_1,\dots,\,q_k)$ satisfying the $(k-1)$ divisibility conditions of Proposition $1$, that uses SWAP gates and controlled SWAP gates. The number of such partitions of $4^n$ allowing for a quantum implementation of the generalized quantum baker map obeys a recursion formula, like provided in Proposition $2$ below. 

\begin{Prop}
The number $Q_n$ of partitions allowing for quantum implementation of a generalized quantum baker map on a square of size $4^n\times 4^n$ satisfies to the recursion:
$$\left\lb\begin{array}{l}
Q_n\;=\;Q_{n-1}^4+1\\
Q_0\;=\;1
\end{array}\right.$$
\end{Prop}

\textsc{Proof.} Using the same notation as in Definition $1$ and additionally setting $N:=4^n$, it suffices to notice that:
\begin{equation}N_{i-1}\leq\frac{N}{4}\Longrightarrow N_i\leq\frac{N}{4}.\end{equation}
Indeed, from $$\left\lb\begin{array}{l}4^{q_i}|N_{i-1}\\4^{q_i}|\frac{N}{4}\end{array},\right.$$
we derive $$4^{q_i}|\frac{N}{4}-N_{i-1}.$$
Hence, $$\frac{N}{4}-N_{i-1}\geq 4^{q_i},$$
that is $$N_i\leq \frac{N}{4}.$$
Given Implication (15), in order to partition $4^n$ in all possible ways, it suffices to partition four segments each of length $4^{n-1}$ independently and then glue the different partitions together. There are $Q_{n-1}^4$ ways of doing so. \hfill $\square$

The following table shows how $Q_n$ grows with the small values of $n$. In this table, we also added a row which  displays the growth of the number of partitions $P_n$ of the quantum baker map on a square of size $2^n\times 2^n$. This is so as to compare the respective key spaces of the scheme with qubits and of the one using ququarts instead. 
\begin{center}
\begin{tabular}{|c|c|c|c|c|c|c|c|c|c|}
\hline
n&2&3&4&5&6&7&8&9\\\hline
$Q_n$&17&83\,522&$>2^{64}$&$>2^{260}$&$>2^{1046}$&$>2^{4184}$&$>2^{16742}$&$>2^{66968}$\\\hline
$P_n$&5&26&677&$458\,330$&$>2^{37}$&$>2^{75}$&$>2^{150}$&$>2^{300}$\\\hline
\end{tabular}
\end{center}
\hfill \textit{Table 1.}

With the generalized quantum baker map used for the scrambling of the pixel positions and of the images, we are now in a position to expose a possible quantum encryption/decryption scheme for a color multi-image, using ququarts. This is the purpose of the next section. 

\subsection{Quantum encryption/decryption scheme}
Given $T$ blocks of $4^n$ images of respective size $2^n\times 2^n$, with possibly $4^{\lT}-T$ blocks filled with blank images, the scrambling stage is simply performed on the color quantum multi-image previously put in QBQQPCMIR representation (9) by using a block and quart plane dependent generalized quantum baker map (GQBM) whose parameters (including the iteration parameter) are secret keys. For each block and quart plane, the GQBM with parameters depending on the block and on the quart plane, acts on the two $n$-ququarts respectively representing the pixel positions on one hand and the images on the other hand: 
\begin{eqnarray}
|M^{'}>&=&\frac{1}{2^{2n+\lT}}\sum_{m,z=0}^{4^n-1}\sum_{t=0}^{4^{\lT}-1}|t>|q>GQBM_{t,q}^{r(t,q)}(|mz>)|RGB>\notag\\
&=&\frac{1}{2^{2n+\lT}}\sum_{m,z=0}^{4^n-1}\sum_{t=0}^{4^{\lT}-1}|t>|q>|mz>|R^{'}G^{'}B^{'}>
\end{eqnarray}
From there, the scrambled color multi-image is ready to get diffused. 

In \cite{YAN}, the authors introduce a 7D hyperchaotic system which reads as follows:
$$\left\lb\begin{array}{l}
\dot{x_1}=a(x_2-x_1)+x_4+bx_6\\
\dot{x_2}=cx_1-x_2-x_1x_3+x_5\\
\dot{x_3}=-dx_3+x_1x_2\\
\dot{x_4}=ex_4-x_1x_3\\
\dot{x_5}=-fx_2+x_6\\
\dot{x_6}=gx_1+hx_2\\
\dot{x_7}=lx_7+mx_4
\end{array}\right.$$
When $(a,b,c,d,e,g,h,l,m)=(10,1,28,8/3,2,1,2,1,1)$ and $f\in (6.3,15.3)$, the system has five positive Lyapunov exponents and the key space of the parameters is infinite. Hyperchaotic systems with more than five dimensions are still quite rare in literature. Hyperchaos was first introduced by Rossler because their dynamic behavior is much harder to predict. A positive Lyapunov exponent means a chaotic behavior and a larger exponent corresponds to a higher sensitivity to initial conditions. A natural way to increase a Lyapunov is following \cite{HZZ} to use a sine chaotification of the original discrete dynamical system. We will make the parameters of the sine chaotification dependent on the blocks. Namely for each block, we pick different parameters. These parameters will be part of the secret keys. The sine chaotification of the original system now reads:
$$\left\lb\begin{array}{l}
\dot{x_1}=sin\big(\pi\la_1^{(t)}(10(x_2-x_1)+x_4+x_6)\big)\\
\dot{x_2}=sin\big(\pi\la_2^{(t)}(28x_1-x_2-x_1x_3+x_5)\big)\\
\dot{x_3}=sin\big(\pi\la_3^{(t)}(-8/3x_3+x_1x_2)\big)\\
\dot{x_4}=sin\big(\pi\la_4^{(t)}(2x_4-x_1x_3)\big)\\
\dot{x_5}=sin\big(\pi\la_5^{(t)}(-7.2x_2+x_6)\big)\\
\dot{x_6}=sin\big(\pi\la_6^{(t)}(x_1+2x_2)\big)\\
\dot{x_7}=sin\big(\pi\la_7^{(t)}(x_7+x_4)\big)
\end{array}\right.$$
The initial conditions are set so that they depend on the plaintext multi-image. It makes the scheme robust to chosen plaintext attacks. For instance, we take $x_1(0)$ to be the total intensity of all the images:
\begin{equation}x_1(0)=\frac{\sum_{t,q,m,z}\big(Q_{t,q,m,z}^{R}+Q_{t,q,m,z}^{G}+Q_{t,q,m,z}^{B}\big)}{4^{2n+\lT}},\end{equation}
where $Q^{C}_{t,q,m,z}$ denotes the quart value of the image $m$ belonging to block $t$ at pixel position $z$ for the color $C\in\lb R,G,B\rb$. Let $T_k$ denote the $k$-th Chebyshev polynomial, namely the unique polynomial of $\mathbb{R}[X]$ such that:
$$\forall \theta\in\mathbb{R},\,T_k(cos \theta)=cos(k\theta)$$
The $T_k$'s can be defined inductively by:
$$\left\lbrace\begin{array}{l}
T_0=1,\,T_1=X\\
T_{k+2}(X)=2XT_{k+1}(X)-T_k(X)
\end{array}\right.$$
For a given color $C$ and for each pixel position $z$, we compute the sum over all the images and all the blocks of the four quart values and take the average over the pixel positions of all these numbers. By taking the integer part of the result, we obtain an integer $r$ (resp $g$, resp $b$). We proceed similarly by fixing the image and the block and computing the sum over all the pixel positions of the four quart values. We then consider the average over all the images belonging to any block and take the integer part of the result. We obtain three other integers, namely one per color, which we denote by $r^{'},\,g^{'},\,b^{'}$. We now set:
\begin{eqnarray}
x_2(0)&=&T_r(x_1(0))\\
x_3(0)&=&T_g(x_1(0))\\
x_4(0)&=&T_b(x_1(0))\\
x_5(0)&=&T_{r^{'}}(x_1(0))\\
x_6(0)&=&T_{g^{'}}(x_1(0))\\
x_7(0)&=&T_{b^{'}}(x_1(0))
\end{eqnarray} 
From the seed, iterate the discrete dynamical system, so as to get six sequences of distinct numbers, and where we ignore the first hundred iterations so as to avoid transcient effects:
\begin{eqnarray} \xi_0,\,\dots,\,\xi_{4^n-1}\\
\io_0,\,\dots,\,\io_{4^n-1}\\
\zeta_0,\,\dots,\,\zeta_{4^n-1}\\\notag \\
\al_0,\,\dots,\,\al_{4^n-1}\\
\be_0,\,\dots,\,\be_{4^n-1}\\
\ga_0,\,\dots,\,\ga_{4^n-1}
\end{eqnarray}
Next, we sort these sequences in increasing order and to a pixel position $z$ (resp an image $m$), we associate the index position $\si^{(r)}_z$ (resp $\ta^{(r)}_m$) of $\xi_z$ (resp $\al_m$) in the sorted sequence and where $(r)$ stands for red. We proceed similarly with the sequences $(18)$ and $(21)$ (resp $(19)$ and $(22)$) associated with color blue (resp green). For an image $m$ belonging to a block $t$ and a pixel position $z$, we compute:
\begin{eqnarray}
\lfloor T_{\si_z^{(r)}}(\be_{4^n-1-z})T_{\ta_m^{(r)}}(\ze_{4^n-1-m}).10^{10}\rf\;\text{mod}\;4^4,\\
\lfloor T_{\si_z^{(g)}}(\ga_{4^n-1-z})T_{\ta_m^{(g)}}(\xi_{4^n-1-m}).10^{10}\rf \;\text{mod}\;4^4,\\
\lfloor T_{\si_z^{(b)}}(\al_{4^n-1-z})T_{\ta_m^{(b)}}(\io_{4^n-1-m}).10^{10}\rf \;\text{mod}\;4^4.
\end{eqnarray}
We derive three $4$-ququart secret keys:
$$\begin{array}{l}
|K_{3,t,m,z}K_{2,t,m,z}K_{1,t,m,z}K_{0,t,m,z}>,\\\\
|L_{3,t,m,z}L_{2,t,m,z}L_{1,t,m,z}L_{0,t,m,z}>,\\\\
|M_{3,t,m,z}M_{2,t,m,z}M_{1,t,m,z}M_{0,t,m,z}>.
\end{array}$$
Let $A$ be the $16\times 16$ unitary matrix which takes a 2-ququarts $|ab>$ and returns:
$$|ab>\;\overset{A}{\lra}\; |a,(a+b)\;\text{mod}\,4>$$
The matrix $A$ gets displayed below. 
\begin{center}
\epsfig{file=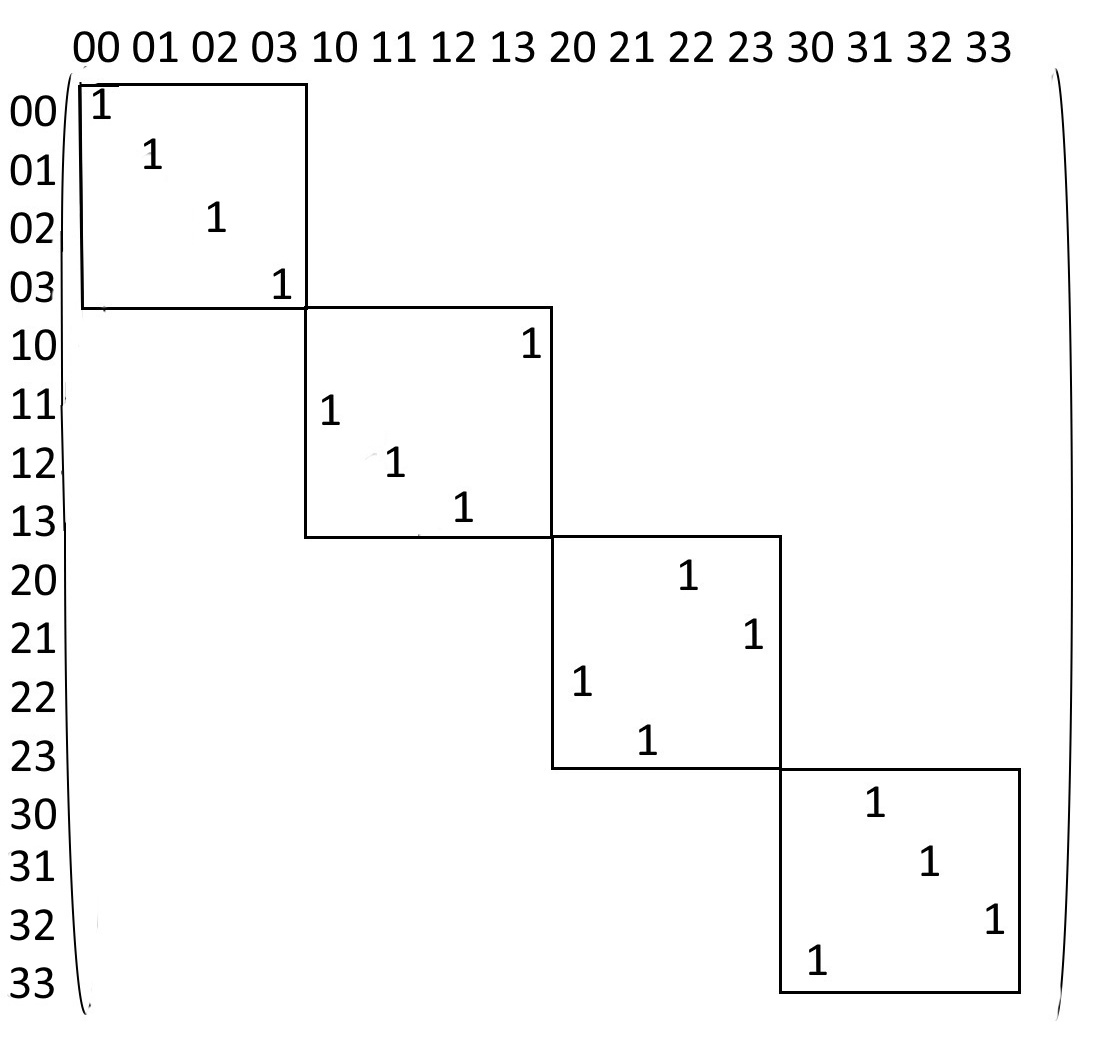, height=8cm}
\end{center}
For the diffusion stage, it will suffice to prepare $4^{2n+1+\lT}$ $3$-ququart ancillas encoding the secret quarts and perform for each $(q,t,m,z)$ three controlled A gates with controls on the quart plane $q$, the block $t$, the image position $m$ and the pixel position $z$:
$$\begin{array}{cc}
C_{q,t,m,z}A&\big(\;|\,K_{q,t,m,z}\,R>\big)\\
C_{q,t,m,z}A&\big(\;|\,L_{q,t,m,z}\,G>\big)\\
C_{q,t,m,z}A&\big(\;|\,M_{q,t,m,z}\,B>\big)
\end{array}$$
Once the color multi-image is scrambled and diffused, it is converted to its classical ciphertext multi-image by projective measurement and transmitted to the receiver. In turn, the receiver puts the ciphertext multi-image into quantum representation and uses the same secret keys previously communicated by the sender, in order to perform the inverse quantum gates in the reverse order. Finally, the receiver puts the decrypted quantum multi-image into classical form by projective measurement in order to be able to read the images. The blank images get discarded. 

The next section addresses a generalisation of the methods presented in $\S\,2$ to quqits. First, we will introduce a space-filling curve in dimension $d$. Second, we will show how it can be used in quantum multi-image encryption. Third, we will generalize the quantum baker map so that it operates on quqits. 

\section{A generalization to quqits with $q$ power of $2$}
\subsection{Space-filling curve in dimension $d$}
\begin{Theorem}
Let $f_1,\dots,f_d:\R\lra\R$ be $d$ continuous $1$-periodic functions such that:
$$\left\lb\begin{array}{cc}
\forall u\in\big[0,\frac{1}{2^{d+1}}\big],& f_1(u)=f_2(u)=\dots = f_d(u)=0\\\\
\forall u\in\big[\frac{2}{2^{d+1}},\frac{3}{2^{d+1}}\big],&f_1(u)=\dots=f_{d-1}(u)=0,\,f_d(u)=1\\\\
&\vdots\\\\
\forall u\in\big[\frac{2^{d+1}-4}{2^{d+1}},\frac{2^{d+1}-3}{2^{d+1}}\big],&f_1(u)=\dots=f_{d-1}(u)=1,\,f_d(u)=0\\\\
\forall u\in\big[\frac{2^{d+1}-2}{2^{d+1}},\frac{2^{d+1}-1}{2^{d+1}}\big],& f_1(u)=\dots =f_d(u)=1
\end{array}.\right.$$
For each $i\in\lb 1,\,\dots,\,d\rb$, define: $$\al_i(u):=\sum_{n=1}^{+\infty}\frac{f_i(2^{(d+1)(n-1)}u)}{2^n}.$$
Then, the map $$U:\begin{array}{ccc}[0,1]&\lra& [0,1]^d\\
u&\longmapsto &(\al_1(u),\,\al_2(u),\,\dots,\,\al_d(u))\end{array}$$
is a space-filling curve. 

\end{Theorem}
\textsc{Proof.} By the same arguments of uniform convergence for the series defining the $\al_i$'s as in the case $d=2$, the map $U$ is continuous. We show that $U$ is surjective. 

Let $u\in [0,1]$. We decompose $u$ in base $2^{d+1}$ as:
$$u=\sum_{n=1}^{+\infty}\frac{u_n}{(2^{d+1})^n},$$
where $u_n\in\lb 0,\,\dots,\, 2^{d+1}-1\rb$. Each $u_n$ is a pit with $p=2^{d+1}$. 
We have:
$$\forall n\geq 1,\;p^{n-1}u=N_n+\frac{u_n}{p}+R_n,$$
with $$N_n\in\N\;\text{and}\;R_n=\sum_{i=n+1}^{+\infty}\frac{u_i}{p^{i+1-n}}\in \Big[0,\frac{1}{p}\Big].$$
Then, since the $f_i$'s are $1$-periodic, we get:
$$\forall i\in\lb 1,\,\dots,\,d\rb,\,\forall n\geq 1,\;2^{-n}f_i(p^{n-1}u)=2^{-n}f_i\bigg(\frac{u_n}{p}+R_n\bigg),\;\text{with $R_n\in \bigg[0,\frac{1}{p}\bigg]$.}$$

Given any $(\be_1,\,\dots,\,\be_d)\in [0,1]^d$, decompose uniquely the $\be_i$'s with $1\leq i\leq d$ as 
$$\be_i=\sum_{n=1}^{+\infty}\frac{\be_{i,n}}{2^n},$$
where the $\be_{i,n}$ are all bits, and using the proper expansion in the case when $\be_i$ admits exactly two such decompositions. Then, we construct a pre-image of $(\be_1,\,\dots,\,\be_d)$ under $U$ as follows:

$$\begin{array}{l}\text{- If $\be_{1,n}=\dots=\be_{d,n}=0$, set $u_n:=0$;}\\
\text{- If $\be_{1,n}=\dots=\be_{d-1,n}=0\;\&\;\be_{d,n}=1$, set $u_n:=2$;}\\
\qquad\qquad\vdots\qquad\qquad\qquad\qquad\qquad\qquad\;\;\vdots\\
\text{- If $\be_{1,n}=\dots=\be_{d-1,n}=1\;\&\;\be_{d,n}=0$, set $u_n:=2^{d+1}-4$;}\\
\text{- If $\be_{1,n}=\dots=\be_{d,n}=1$, set $u_n:=2^{d+1}-2$.}\end{array}$$
Let $$u:=\sum_{n=1}^{+\infty}\frac{u_n}{2^{n(d+1)}}.$$
By construction, we have $\al_i(u)=\be_i$ for every integer $i$ with $1\leq i\leq d$. 
Thus, $U$ is a space-filling curve of the unit hypercube of dimension $d$.\hfill $\square$

Given a $d$-dimensional hypercube of size $2^n$ and an integral point $x=(x_1,\dots,x_d)$ of the hypercube, decompose for each coordinate:
$$\frac{x_i}{2^n}=\sum_{k=1}^n\frac{a_{i,n-k}}{2^k},$$
where the $a_{i,l}$'s for $0\leq l\leq n-1$ are uniquely determined by the binary decomposition of the integer $x_i$. Now set $u_{n-k}:=0$ if $a_{1,n-k}=\dots =a_{d,n-k}=0$ (resp $u_{n-k}:=2$ if $a_{1,n-k}=\dots = a_{d-1,n-k}=0$ and $a_{d,n-k}=1$, $\dots$, resp $u_{n-k}:=2^{d+1}-4$ if $a_{1,n-k}=\dots = a_{d-1,n-k}=1$ and $a_{d,n-k}=0$, resp $u_{n-k}:=2^{d+1}-2$ if $a_{1,n-k}=\dots = a_{d,n-k}=1$). Then, 
$$\sum_{k=1}^n\frac{u_{n-k}}{2^{(d+1)k}}$$
is a pre-image of $\frac{x}{2^n}$ under the space-filling curve of Theorem $2$. 
Any such integral point $x$ can be represented using $n$ quqits $|0>$, $|1>$, $\dots$, $|q-1>$, where $q=2^d$, respectively corresponding to $u_i=0$, $u_i=2$, $\dots$, $u_i=2q-2$. From a representation using $dn$ qubits, we derived a representation using $n$ quqits with $q=2^d$. 
\subsection{Quantum baker map on qutits}
Given two $n$-qutits, we shuffle them by using a quantum baker map defined on a discretized square of size $t^n\times t^n$. 
\begin{Definition}
The generalized discrete baker map with parameters $(q_1,\dots, q_k)$ is defined for a square of size $t^n\times t^n$ with
$$t^n=t^{q_1}+\dots +t^{q_k}$$
as
$$\left\lbrace\begin{array}{l}
N_0=0\; \text{and}\; N_i=t^{q_1}+\dots +t^{q_i}\;\text{and}\;
x=N_{i-1},N_{i-1}+1,\dots,N_i-1\\
(x^{'},y^{'})=\Bigg(t^{n-q_i}(x-N_{i-1})+y\,\text{mod}\,t^{n-q_i},N_{i-1}+\frac{y-y\,\text{mod}\,t^{n-q_i}}{t^{n-q_i}}\Bigg)
\end{array}\right.$$
\end{Definition}

\begin{Prop}
(i) The map defined by
$$G_s(x,y)=\Bigg( t^{n-s}x\,\text{mod}\,t^n+y\,\text{mod}\,t^{n-s},\frac{y-y\,\text{mod}\,t^{n-s}}{t^{n-s}}+x-x\,\text{mod}\,t^s\Bigg)$$
has a quantum implementation. \\
(ii) Each subfunction on $\lbrace N_{i-1},\dots,N_i-1\rbrace$ of the generalized discrete baker map is $G_{q_i}$ if and only if $t^{q_i}|t^{q_1}+\dots+t^{q_{i-1}}$.
\end{Prop}
\textsc{Proof.} It is straightforward to adapt the proof of Proposition $1$. \hfill $\square$ 

Again, we have a quantum circuit using SWAP and controlled SWAP gates for implementing the discretized baker map on the square $t^n\times t^n$, whose parameters $(q_1,\,\dots,\,q_k)$ satisfy to the divisibility relations $t^{q_i}|t^{q_1}+\dots +t^{q_{i-1}}$ for each integer $i$ with $2\leq i\leq k$. Moreover, the number of partitions allowing for a quantum implementation of the discretized baker map gets provided in Proposition $4$ below. 

\begin{Prop}
The number $T_n$ of partitions allowing for quantum implementation of a generalized quantum baker map on a square of size $t^n\times t^n$ satisfies to the recursion:
$$\left\lb\begin{array}{l}
T_n\;=\;T_{n-1}^t+1\\
T_0\;=\;1
\end{array}\right.$$
\end{Prop}

For instance, when $n=3$ and $t=8$ this number is $19031147999601100801$, which corresponds to the number of admissible parameters for a quantum implementation of a discretized baker map used for scrambling two $3$-quoctits, versus only $P_3=26$ admissible partitions for scrambling two $3$-qubits. 

\subsection{Application to multi-image quantum encryption with quoctits}

In light of the results from the previous two parts, we discuss below a non-exhaustive list of options for multi-image quantum encryption using quoctits. 

\subsubsection{The three-stage scrambling}
In \cite{LE1}, we used a bit plane quantum representation for a multi-image and formed blocks of images, with each block containing $2^{\lL}$ images, where $L$ denotes the number of bit planes. We then used this new representation to scramble the bit planes and the images belonging to a given block using the quantum baker map on qubits. In the usual case of $8$ bit planes, the quantum baker map gets applied on a square of size $2^3\times 2^3$ and by Table 1, there are $P_3=26$ ways of choosing the quantum baker map. Suppose we have eight bit planes, eight predefined colors with intensity ranging from $0$ to $255$ and blocks of eight images of standard size $512\times 512$. By $\S\,3.1$, the eight bit planes, eight colors and eight images can be represented using a $3$-quoctit. An  horizontal pixel coordinate requires the use of a $3$-quoctit and so does a vertical pixel coordinate. We can then realize a three-stage scrambling by performing a quantum baker map on the $3$-quoctit encoding the horizontal pixel coordinate and the $3$-quoctit encoding the triple (bit plane, color, image); performing a quantum baker map on the               
$3$-quoctit encoding the vertical pixel coordinate and the $3$-quoctit encoding the triple (bit plane, color, image); performing a quantum baker map on the two $3$-quoctits encoding the pixel positions; all these stages in any given order. 

The quantum representation uses a total of $10+\lTe$ quoctits where $T$ denotes the number of blocks. For instance, it uses $12$ quoctits if we encode a total of $512$ images. 
In comparison, the quantum representation of \cite{LE1}, generalized to a color multi-image within the same color system, uses $25+\lM$ qubits, where $M$ denotes the total number of images to encrypt, that is $34$ qubits in the case when $M=512$ images. In the latter case, the storage space is divided by a factor of almost three. More generally, for $(2^3)^k$ images to encrypt, the scheme with quoctits uses $9+k$ quoctits while the scheme with qubits uses $25+3k$ qubits. The ratio is still roughly three and is independent from how the number of images to encrypt grows. 

An important feature of this new scheme is the following. In \cite{LE1}, we achieve a pixel position independent scrambling of the bit planes and the images. It means that at each pixel position, a scrambling of the bit planes and of the images is realized with admissible parameters for the quantum baker map depending on the pixel position. This is impractical as a total of $2^{18}$ secret keys for the simple scrambling of the images and of the bit planes must be previously and independently transmitted to the receiver. However, the idea arose from the fact that there are only $26$ possible quantum baker maps for scrambling a square of size $2^3\times 2^3$. By Table $4$ of \cite{BA0}, the scrambling period for the quantum baker map on such a square is $840$ (versus only $6$ and $12$ for the Arnold quantum transform and the Fibonacci quantum transform respectively). The period on the whole square is calculated from the least common multiple of all the discretized coordinates'periods. Then, a remedy consists of iterating the quantum baker map, instead of proceeding with the independent scrambling. The iteration parameter is one of the secret keys. This now offers $21840$ possibilities, instead of $26$. In the present scheme, there are $19031147999601100801$ ways of scrambling two $3$-quoctits, hence there is no need to iterate the quantum baker map. 

Once the color multi-image is scrambled, it gets diffused using a sine chaotification of a $4D$ Lorenz hyperchaotic system. The generalization of the Lorenz $3D$ chaotic system to a $4D$ hyperchaotic system was achieved by the authors of \cite{WAN}. Their system is as follows. 
$$\left\lb\begin{array}{l} \dot{x}=a(y-x)+w\\
\dot{y}=cx-y-xz\\
\dot{z}=xy-bz\\
\dot{w}=-yz+rw
\end{array}\right.$$
When $a=10$, $b=8/3$, $c=28$ and $r\in (-1.52,-0.06]$, the system enters a hyper-chaotic state. We note that the hyperchaotic system of \cite{YAN} used in our ququart model is a generalization of this $4D$ system to the higher dimension $7D$.  We now consider a sine chaotification of this system, where the parameters of the sine chaotification depend on the block, and where we set $r:=1.1$. 
$$\left\lb\begin{array}{l} \dot{x}=sin\big(\pi\mu_1^{(t)}\big(10(y-x)+w\big)\big)\\
\dot{y}=sin\big(\pi\mu_2^{(t)}\big(28x-y-xz\big)\big)\\
\dot{z}=sin\big(\pi\mu_3^{(t)}\big(xy-8/3z\big)\big)\\
\dot{w}=sin\big(\pi\mu_4^{(t)}\big(-yz-1.1w\big)\big)
\end{array}\right.$$
Choose an initial seed that depends on the plaintext images to avoid CPA attacks. 
For instance, $x(0)$ is the intensity of all the images:
\begin{equation}x(0):=\sum\frac{\text{all bit values}}{2^{27+3\lTe}}.\end{equation}
Next, let $s$ denote the sum of all the bit values and take 
\begin{equation} y(0):= T_s(x(0)),\end{equation} where $T_s$ is the $s$-th Chebyshev polynomial. Further, for each pixel position, compute the sum of all the bit values at that pixel position. Then take the integer part of the average of these numbers over all the pixel positions. Get an integer $s^{'}$ and set 
\begin{equation}z(0):=T_{s^{'}}(x(0)).\end{equation}
Last, for each triple consisting of an image, a bit plane and a color, and for each block, compute the sum of all the bit values over the pixel positions and then take the integer part of the average of these numbers so as to obtain an integer $s^{''}$. Then define
\begin{equation}w(0):=T_{s^{''}}(x(0)).\end{equation}
Iterate the discrete dynamical system from this seed as long as is necessary to obtain three sequences of $2^9$ distinct numbers and during the process ignore the first hundred iterations to avoid transient effects. Thus obtain: 
\begin{equation}x_0,\,\dots,\,x_{2^9-1},\end{equation}
\begin{equation}y_0,\,\dots,\,y_{2^9-1},\end{equation}
\begin{equation}z_0,\,\dots,\,z_{2^9-1}.\end{equation}
Sort these three sequences by increasing order and to each horizontal (resp vertical) pixel coordinate $i$ (resp $j$), associate the index position $n_i$ (resp $m_j$) of $x_i$ (resp $y_j$) in the sequence (37) (resp (38)); to each triple (image, bit plane, color) $l$, represented by its $3$-quoctit $|l>$, associate the index position $k_l$ of $z_l$ in the sequence (39). Then compute the secret bit:
\begin{equation}
\lfloor T_{n_i}(y_j)T_{m_j}(z_{l})T_{k_l}(x_i).10^5\rfloor\pmod 2
\end{equation}
Prepare ancilla quoctits which encode the secret bits and perform controlled quantum gates with controls on the block, the pixel position, the triple (image, bit plane, color) realizing the addition modulo 2 on the quoctit representing the bit value and the quoctit representing the secret bit, while leaving the secret ancilla quoctit invariant. 
\subsubsection{The mixed scrambling}
In this part, given an even number $2k$ of quoctits in the quantum representation of the multi-image, we divide the quoctits into two groups of equal size $k$ and shuffle the resulting two $k$-quoctits with a quantum baker map, independently from what the quoctits represent and under the only condition that all the quoctits involved appear within the full range of possibilities as a superposition in the quantum representation. The latter condition allows for instance that all the pixel positions are still represented in the scrambled multi-image. 

We describe below several possible situations of application, corresponding to a wide range of possibilities.  

a) Suppose we have blocks of $512$ color images of size $2^9\times 2^9$ and eight bit planes, where we use the RGB system and the bit plane representation model. We use a $9$-quoctit to represent the pixel positions and the images and one quoctit to represent the bit planes. We have a total of ten quoctits (excluding those representing the blocks and those three representing bit values of R, G and B respectively),  which we split into two arbitrary groups, each containing five quoctits. We then scramble the two $5$-quoctits using the quantum baker map (abbreviated QBM) of Definition $2$, where $t=8$ and $n=5$. There are over $10^{152}$ admissible parameters for the QBM. The quantum representation uses a total of $13+\lTe$ quoctits, where $T$ is the number of blocks. In comparison, the scheme of \cite{LE1}, applied in the same color setting, uses $33+\lTd$ qubits. 

For the diffusion, use the same hyperchaotic system as described in $\S\,3.3.1$, whose parameters depend on the block. For initial conditions, still take for $x(0)$ the total intensity of all the images. For $y(0)$, apply a Chebyshev transform with index the sum of all the bit values of the three colors. The next two indices of Eq. (35) and Eq. (36) are computed using similar ideas as before and adapting them, namely, for a pixel position and an image (resp for a block and a bit plane), compute the sum over all the blocks and all the bit planes (resp over all the $9$-quoctits encoding a pixel position and an image) of all the bit values on the three colors, then take the integer part of the average of all these numbers. Iterate the dynamical system so as to obtain three sequences of $2^{27}$ distinct numbers $$x_0,\dots,x_{2^{27}-1},y_0,\dots,y_{2^{27}-1}, z_0,\dots,z_{2^{27}-1}.$$ Sort these three sequences by increasing order and for each $9$-quoctit $|k>$ encoding the pixel position and image, denote by $n_k$ (resp $m_k$, resp $r_k$) the index position of $x_k$ (resp $y_k$, resp $z_k$) in the sorted sequence. Then, compute 
\begin{equation}\lf T_{n_k}(y_k)T_{m_k}(z_k)T_{r_k}(x_k).10^{20}\rf\pmod{2^{24}},\end{equation}
and derive eight secret bits per color R, G or B. For each color C, each bit plane, each quoctit $|k>$ encoding the pixel position and the image and each block, use a quoctit ancilla encoding the corresponding secret bit and perform a controlled gate with control on the block, the quoctit $|k>$ and the bit plane, which realizes the addition modulo two of the secret bit and of the bit value of color C, while leaving the secret quoctit unchanged. 

b) Suppose we have blocks of $256$ images of size $2^8\times 2^8$, eight predefined colors, defined by their RGB codes. These eight colors could for instance be red, yellow, orange, blue, green, purple, black and pink. There are eight bit planes, encoding the intensity of each color. The pixel positions and images can be encoded using an $8$-quoctit, the colors can be encoded using one quoctit, and so do the bit planes. It yields a total of ten quoctits to be split into two groups of five quoctits each, that are ready to get scrambled under a quantum baker map. The quantum representation requires a total of $11+\lTe$ quoctits, with $T$ the number of blocks. If $T=8^k$ a scheme like in \cite{LE1} with these same modalities requires $31+3k$ qubits. 

For the diffusion stage, we adapt the diffusion scheme of a). For each of the eight colors, one could pick a different control parameter $r$ in the original $4D$ hyperchaotic system. For the choice of integer $s^{'}$, one should sum over all the colors, all the bit planes and all the blocks all the bit values for a given $8$-quoctit and then average over all the $8$-quoctits. For the integer $s^{''}$, proceed similarly by fixing the color, the bit plane and the block and summing over all the $8$-quoctits the bit values, then take the integer part of the average of all the so-computed integers. The three pseudo-random sequences should now have length $2^{24}$ and for each $8$-quoctit $|k>$ representing an image and a pixel position, one should compute instead of (41):

\begin{equation}\lf T_{n_k}(y_k)T_{m_k}(z_k)T_{r_k}(x_k).10^{10}\rf\pmod{2^{8}},\end{equation}
in order to derive eight secret bit values to be XORed with those from the eight bit planes. 

c) Suppose we have blocks of $16$ images of size $512\times 512$ encoded with $16$ colors such as black, yellow, gold, red, burgundy, orange, pink, purple, violet, green, olive, lime, blue, cyan, beige, brown, and $16$ bit planes for encoding the intensity of each color. We have $6$ quoctits to represent the pixel positions and $4$ quoctits to represent the images, the colors and the bit planes. Hence a total of $10$ quoctits which we split into two groups of five quoctits. We then scramble the two $5$-quoctits using a quantum baker map. The quantum representation uses a total of $11+\lTe$ quoctits, with $T$ the number of blocks. A scheme like in \cite{LE1}, generalized to colors and under the same data, requests $31+\lTd$ qubits. 

This time, the diffusion scheme is identical to the one used in $\S\,3.3.1$, except a triple (image, bit plane, color) is represented by a $4$-quoctit, instead of a $3$-quoctit. Consequently, we need a pseudo-random sequence of length $2^{12}$ instead of $2^9$ for the $z$ coordinate of the discrete dynamical system. The rest is unchanged.

\subsubsection{Some variants}

Last, we present some variants of the scrambling schemes presented above. In what follows, we only focus on the scrambling part and leave the diffusion stage to the reader. 

$\alpha)$ Our first variant uses a combination of mixed scrambling and three stage scrambling and allows to encode slightly over a million images all at once. Namely, suppose we have blocks of $256$ images of size $256\times 256$. We use an $8$-quoctit to represent the pixel positions and the images. The mixed scrambling part consists of dividing this group of $8$ quoctits into two subgroups of $4$ quoctits. Suppose now we have $8^4$ blocks of images and the blocks can be represented using a third $4$-quoctit. We now have three $4$-quoctits at our disposal and apply to it a three-stage scrambling, following the same modalities as in $\S\,3.3.1$. 
Yet an alternative is to do a simple mixed scrambling on the $12$-quoctits after splitting them into two $6$-quoctits, following the same general guideline as in $\S\,3.3.2$. The quantum representation uses a total of $16$ quoctits, where we work with eight bit planes and the RGB color system. In contrast, the representation of \cite{LE1} generalized to colors, uses a total of $42$ qubits. 

$\beta)$ Our second variant assumes that we have a total of $4096$ color images of size $512\times 512$ to encrypt. We can form eight blocks of $8^3$ images. The images are represented using eight colors and eight bit planes. Then we have a $3$-quoctit representing the color, the block and the bit plane. And we have another $3$-quoctit representing the images. We scramble both $3$-quoctits with a quantum baker map. The quantum representation uses a total of $13$ quoctits, instead of $37$ qubits in a \cite{LE1} type scheme. 

Alternatively, we can form $256$ blocks of $16$ images. In this scheme, we use $16$ colors such as black, yellow, gold, red, burgundy, orange, pink, purple, violet, green, olive, lime, blue, cyan, beige, brown, and the images all have size $256\times 256$. We also use $16$ bit planes encoding the intensity of each color. It yields a $4$-quoctit for the images, the colors and the bit planes and an $8$-quoctit for the pixel positions and the blocks, hence a total of $12$ quoctits to be split into two groups of $6$ quoctits. We then realize a mixed scrambling using a quantum baker map on the square $8^6\times 8^6$. This scrambling scheme requests a quantum representation using a total of $13$ quoctits,  versus $37$ qubits for the equivalent scheme of \cite{LE1}. 

\subsubsection{The monster scheme}
In this scheme, we consider $256$ blocks, each containing $16\,777\,216$ images of size $256\times 256$. We use an $8$-quoctit to encode the pixel positions and the blocks and an $8$-quoctit to encode the image position inside a block. We then scramble the two $8$-quoctit using the quantum baker map. The total number of quoctit needed is $20$ quoctits. In comparision, the corresponding scheme of \cite{LE1} uses a total of $54$ qubits. 

For the diffusion stage, consider for instance a sine chaotification of the $4D$ hyperchaotic system previously defined in $\S\,3.3.1$ with parameters $\mu_1,\,\mu_2,\,\mu_3,\,\mu_4$ chosen independently for each color red, blue and green. For initial conditions, take $x(0)$ to be the total intensity of all the images, for $y(0)$ the total intensity for color $C\in\lb R,G,B\rb$. For a given pixel position and block, compute the sum of all the bit values for color $C$ of all the images belonging to that block, and then take the integer part $s$ of the average of these numbers over all the pixel positions and blocks. For a given image position, compute the sum of all the bit values for color C for all the pixel positions of all the images at this position of all the blocks. Then take the integer part $s^{'}$ of the average of all these numbers. Then set:
$$\left\lb\begin{array}{l} z(0):=T_s(x(0)),\\
w(0):=T_{s^{'}}(x(0)).\end{array}\right.$$
For a given color C, iterate the so-defined discrete dynamical system in order to get two pseudo-random sequences of $2^{24}$ distinct numbers:
$$\begin{array}{l}
x_0,\,\dots,x_{2^{24}-1}\\
y_0,\,\dots,\,y_{2^{24}-1}
\end{array}$$
Sort these two sequences in increasing order, and for a given $8$-quoctit $|k>$ representing a triple (horizontal pixel coordinate, vertical pixel coordinate, block), assign the index position $n_k$ of $x_k$ in the first sorted sequence. For each image position $m$, assign the index position $r_m$ of $y_m$ in the second sorted sequence. 
Then compute for each such $m$ and $|k>$ 
$$\lfloor T_{n_k}(y_{2^{24}-k})T_{r_m}(x_{2^{24}-m})\rfloor\;\pmod{2^8}.$$
Derive a secret $8$-qubit for color C. For each bit plane, image position and $8$-quoctit $|k>$, prepare a secret $3$-quoctit ancilla encoding the $3$-qubit of the three secret bits, one per color, and perform a controlled gate with control the bit plane, the image position, $|k>$, realizing the three additions modulo $2$, one per color, of each secret bit and bit value, while leaving the secret $3$-quoctit invariant. 

This scheme allows to encrypt a total of $4.\,294.967.296$ images, that is pretty much half the size of the world population. We could for instance imagine steganography to hide some text information in each portrait image. 

\newpage\pagestyle{empty}

\subsubsection{Quantum encryption schemes with quoctits library}
\hspace{-5.5cm}
\begin{tabular}{r|c|c|c|c|c|c|l}
\hline
$\#$ images&$\begin{array}{l} T\,\#\text{blocks}\\\text{8 per block}\end{array}$&$\begin{array}{l} T\,\#\text{blocks}\\\text{512/block}\end{array}$&$\begin{array}{l} T\,\#\text{blocks}\\\text{256/block}\end{array}$&$\begin{array}{l} T\,\#\text{blocks}\\\text{16/block}\end{array}$&\begin{tabular}{c}Up to $2^{20}$\\$\nts>1$ million\\256/block\end{tabular}&\begin{tabular}{c}Up to 4096\\512/block\\16/block\end{tabular}&\begin{tabular}{c}\nts\nts\nts\nts\nts Up to\\\nts\nts\nts\!\!\nts $2^{32} (>\!4$\\\nts\nts\nts\nts\!billions\!)\end{tabular}\\\hline
$\begin{array}{l}
\text{Images}\\
\text{size}
\end{array}$&$512\times 512$&$512\times 512$&$256\times 256$&$512\times 512$&$256\times 256$&$512\times 512$&$\nts 256^2$\\\hline
$\;\begin{array}{l}\#\;\text{blocks,}\\
\text{eventually}\\\text{filled}\\
\text{with}\\\text{blank}\\\text{images}\end{array}$&$8^{\lTe}$&$8^{\lTe}$&$8^{\lTe}$&$8^{\lTe}$&$2^{12}$&\begin{tabular}{c}8\\256\end{tabular}&$2^8$\\\hline
$\begin{array}{l}\text{Blank}\\
\text{images}\end{array}$&Yes&Yes&Yes&Yes&Possibly&Possibly&$\nts\!\!$P\!ossibly\\\hline
$\begin{array}{l}\text{Color}\\\text{system}\end{array}$&8&RGB&8&16&RGB&8&\nts RGB\\\hline
$\begin{array}{l}\#\;\text{bit}\\\text{values}\\
\text{in}\\\text{BRQMI}\end{array}$ &1&3&1&1&3&1&3\\\hline
$\begin{array}{l}\#\\\text{bit}\\\text{planes}\end{array}$&8&8&8&16&8&\begin{tabular}{c}8\\16\end{tabular}&8\\\hline
$\begin{array}{l}\text{Key data}\\\text{for}\\
\text{quantum}\\
\text{rep.}\end{array}$&\begin{tabular}{l}\begin{tabular}{l}3-quoctit\\x p.c.\end{tabular}\\\begin{tabular}{l}3-quoctit\\y p.c.\end{tabular}\\\begin{tabular}{l}3-quoctit\\$\left|\begin{array}{l}\text{colors}\\\text{images}\\\text{bit planes}\end{array}\right.$\end{tabular}\end{tabular}&\begin{tabular}{c}\begin{tabular}{c}9-quoctit\\$\left|\begin{array}{l}\text{x p.c.}\\\text{y p.c.}\\\text{images}\end{array}\right.$\end{tabular}\\\begin{tabular}{c}1-quoctit\\bit planes.\end{tabular}\end{tabular}&\begin{tabular}{c}\begin{tabular}{c}8-quoctit\\$\left|\begin{array}{l}\text{x p.c.}\\\text{y p.c.}\\\text{images}\end{array}\right.$\end{tabular}\\\begin{tabular}{c} 1-quoctit\\colors\end{tabular}\\\begin{tabular}{c}1-quoctit\\bit planes\end{tabular}\end{tabular}&\begin{tabular}{l}\begin{tabular}{l}6-quoctit\\pixel pos.\end{tabular}\\\begin{tabular}{l}4-quoctit\\$\left|\begin{array}{l}\text{colors}\\\text{images}\\\text{b.p.}\end{array}\right.$\end{tabular}\end{tabular}&\begin{tabular}{c}\begin{tabular}{c}8-quoctit\\$\left|\begin{array}{l}\text{x p.c.}\\\text{y p.c.}\\\text{images}\end{array}\right.$\end{tabular}\\\begin{tabular}{c}4-quoctit\\blocks\end{tabular}\end{tabular}&\begin{tabular}{l}\begin{tabular}{l}3-quoctit\\$\left|\begin{array}{l}\text{blocks}\\\text{colors}\\\text{b.p.}\end{array}\right.$\end{tabular}\\\begin{tabular}{l}3-quoctit\\images\end{tabular}\\\hline\begin{tabular}{l}4-quoctit\\$\left|\begin{array}{l}\text{images}\\\text{colors}\\\text{b.p.}\end{array}\right.$\end{tabular}\\\begin{tabular}{l}8-quoctit\\$\left|\begin{array}{l}\text{blocks}\\\text{x p.c}\\\text{y p.c.}\end{array}\right.$\end{tabular}\end{tabular}&$\nts\nts\nts\nts\nts\nts$\begin{tabular}{l}\begin{tabular}{l}\!8quoctit\\$\left|\begin{array}{l}\text{x p.c.}\\\text{y p.c.}\\\!\text{blocks}\end{array}\right.$\end{tabular}\\\begin{tabular}{l}\!8quoctit\\\!images\end{tabular}\end{tabular}
\\\hline
$\#$ quoctits&\nts\nts\begin{tabular}{c|c}10&\\+&13\\$\lTe$&\end{tabular}&\nts\nts\begin{tabular}{c|c}13&\\+&14\\$\lTe$&\end{tabular}&\nts\nts\begin{tabular}{c|c}11&\\+&13\\$\lTe$&\end{tabular}&\nts\nts\begin{tabular}{c|c}11&\\+&14\\$\lTe$&\end{tabular}&16&13&20\\\hline
\begin{tabular}{c}$\#$ qubits\\\cite{LE1}\end{tabular}&\nts\nts\nts\begin{tabular}{c|c}25&\\+&37\\$\lM$&\end{tabular}&\nts\nts\nts\begin{tabular}{c|c}24&\\+&36\\$\lM$&\end{tabular}&\nts\nts\nts\begin{tabular}{c|c}23&\\+&35\\$\lM$&\end{tabular}&\nts\nts\!\!\begin{tabular}{c|c}27&\\+&39\\$\lM$&\end{tabular}&42&37&54\\\hline
$\;\begin{array}{l}\text{Scrambling}\\
\text{type}\end{array}$&three stage&mixed&mixed&mixed&\begin{tabular}{cc}mixed and\\
three stage\\\hline mixed\end{tabular}&\begin{tabular}{c}regular\\\hline mixed \end{tabular}&\nts regular\\\hline
$\;\,\begin{array}{l}\text{Square of} \\
\text{application}\\
\text{of QBM}\end{array}$&$8^3\times 8^3$&$8^5\times 8^5$&$8^5\times 8^5$&$8^5\times 8^5$&\begin{tabular}{c}$8^4\times 8^4$\\$8^6\times 8^6$\end{tabular}&\begin{tabular}{c}$8^3\times 8^3$\\$8^6\times 8^6$\end{tabular}&\nts $8^8\times 8^8$\\\hline
$\;\;\,\begin{array}{l}\#\;\text{choices}
\\\text{parameters}
\end{array}$&$>10^{19}$&$>10^{1233}$&$>10^{1233}$&$>10^{1233}$&\begin{tabular}{c}$>10^{154}$\\$>10^{9871}$\end{tabular}&\begin{tabular}{c}$\nts\nts>10^{19}$\\$>10^{9871}$\end{tabular}&\begin{tabular}{l}$\nts\nts>$\\\nts\nts\nts\nts\!$10^{631748}$\end{tabular}\\\hline
$\;\;\,\;\begin{array}{l}\;\;\;\;\#\text{$\,$C$\!$ gates}\\
\;\;\;\text{$\,$for $\!$diffusion}\end{array}$&$2^{27}$&$2^{30}$&$2^{30}$&$2^{30}$&$2^{27}$&\begin{tabular}{c}$2^{27}$\\$2^{18}$\end{tabular}&$2^{21}$\\\hline
\end{tabular}

\pagestyle{plain} Below are a few explanations and comments regarding the table above. 

In the table above, the numbers inside the multi-columns are obtained by setting $M=4096$ as the total number of images, like in scheme number 6 when moving from left to right inside the table. 

The abbreviations p.c and b.p respectively stand for ``pixel coordinate" and ``bit planes". 

Some of the schemes have a set number of blocks and number of images per block, in order to allow for a certain type of scrambling. In that case, if the number of images to encrypt is less than the upper bound corresponding to the number of blocks times the number of images per block, some blank images must be added, until all the blocks are filled. For some other schemes, the important data is the number of images per block, say $m$, and the number of blocks gets determined from the total number of images to encrypt, which can be any number $M$, by computing $T=\lceil M/m\rceil$. It means that the last block possibly contains some blank images. Moreover, the actual number of blocks in the quantum representation is rather $8^{\lTe}$, a number that is greater than or equal to $T$. It eventually adds some additional blank images. For a random number of images to transmit, chosen by the user, it is thus highly likely that blank images will need to be added. We note that in the first scheme type, the scrambling is identical for all the blocks.
Whereas, in the second scheme type, the blocks are an integrant part of the shuffling process. This induces an even better shuffling. 

In the table, whenever a vertical bar appears below a given $n$-quoctit, it means that the $n$-quoctit is used to represent a point of the cube of size $2^n\times 2^n\times 2^n$, following the definition arising from the space-filling curve introduced in $\S\,3.1$, in the special case when $d=3$. 

Last, we comment on the last row of the table. In order to reduce the circuit complexity without affecting the security of the protocols significantly, we do the following. Whenever it is allowed by the quantum representation of the multi-image (that is in schemes 1--5), and contrary to in our discussion, the numbers appearing in the table correspond to a diffusion scheme that is identically performed on each block. Also, in the first alternative of scheme number 6, we perform the same diffusion for all the images within a block, but the diffusion pattern varies from a block to another block. We note that the circuit complexity in each case is very high. This should actually not be a concern. Indeed, there exist diffusion schemes inducing a lower circuit complexity. This fact is even more relevant to the second alternative of scheme number 6, as well as to scheme number 7. Namely, in the same way as we formerly introduced the concept of ``mixed scrambling", we introduce the concept of ``mixed diffusion". The idea is to diffuse along certain quoctits only, belonging to various data types. For instance, in the second alternative of scheme 6, we could pick $6$ quoctits out of the $12$ quoctits belonging to the key data for quantum representation, with quoctits belonging to each of the two groups. We could then fix 
six quoctits for which the diffusion is achieved independently from their values, while the diffusion is dependent from the octit values taken by the other six quoctits. 
Likewise, in scheme 7, we have $16$ key quoctits at our disposal, and could select six of them so that they belong to both groups of eight quoctits. We then let these six quoctits determine the diffusion for the quantum multi-image. 
This diffusion method leads to the last two values of $2^{18}$ and $2^{21}$ inside the table (in the latter case, XOR operations are achieved for the three colors RGB for each $6$-quoctit and each of the 8 bit planes). 

\subsection{Diffusion and topological quantum computation}
\pagestyle{plain}
We have seen in $\S\,1$ that quqits with $q$ power of two theoretically exist in topological quantum computation with anyons. If $q=2^d$, such a quqit can for instance be realized by four $SU(2)_{2^{d+1}-1}$ anyons of topological charge $2^d-1$, $2^d$, $2^d$, $2^d-1$. 
However, in order to perform universal quantum computation with such anyons, a leakage-free two quqit entangling gate is needed. \\

\textbf{Open problem 1.} Let $q=2^d$. Is it possible to realize a leakage-free $2$-quqit entangling gate on two quqits of respective topological charges $2^d-1$, $2^d$, $2^d$, $2^d-1$ with $SU(2)_{2^{d+1}-1}$ anyons ?\\

An interesting feature of quantum computing with anyons for quantum multi-image encryption/decryption with qubits relates to the diffusion stage. Diffusion is usually achieved by preparing pseudo-random secret bits and XOR-ing them with the bit values of the quantum multi-image. The question about how to perform these XOR operations on a quantum computer got raised in \cite{GRA}. In any general quantum computing approach, these can be achieved by preparing ancilla qubits encoding the secret bits and performing as many controlled CNOT gates as is necessary, depending on the scheme, where the CNOT gate gets applied to the qubit encoding the bit value which serves as target qubit, and to the secret qubit which serves as control qubit. In the topological approach consisting of braiding and measuring $SU(2)_4$ anyons, there exists an alternative approach consisting of preparing an ancilla which is an entangled superposition with the same number of qubits as in the quantum representation of the scrambled multi-image, and where the qubits encoding the bit values have been replaced with qubits encoding the secret bits. It then suffices to perform a single CNOT gate on the qubit encoding the secret bit (control qubit) and the qubit encoding the bit value (target qubit), while fusing two by two adequately all the other qubits belonging to the scrambled quantum multi-image and to the ancilla respectively. At the end of the process, the qubit encoding the secret bits is entangled with the diffused quantum multi-image. This qubit should simply be discarded and never be measured by projective measurement, as it does not carry any information about the ciphertext multi-image and the information it carries is moreover secret. There exists a way to realize the topological fusion of two qubits with $SU(2)_4$ anyons, under the qubit encoding $1221$, see \cite{LEQ}\cite{LEM}. Measurements are a key ingredient and recovery procedures do exist in the case of unsatisfactory measurement outcomes. The technique of topological fusion is in fact not specific to topological qubits. Namely, it can also be achieved with qutrits \cite{LEQ}. Topological fusion of two qutrits is in fact part of the method used in \cite{LEQ} in order to realize an irrational qutrit phase gate on four $SU(2)_4$ anyons of topological charge $2$. This important result leads to single universality with qutrits when quantum computing with $SU(2)_4$ anyons \cite{LEQ}. Thus, it is legitimate to ask whether topological fusion could be generalized to qudits, within higher levels of the theory. This problem is stated below in our special case of interest. \\

\textbf{Open problem 2.} Let $q=2^d$. With $SU(2)_{2^{d+1}-1}$ anyons, is it possible to achieve the topological fusion of two quqits, respectively formed by four anyons of topological 
charges $2^d-1$, $2^d$, $2^d$, $2^d-1$, where the second quqit is a certain type of ancilla to be determined, allowing for the fusion ?\\

We note that both Problems 1 and 2 could be reformulated in terms of quqits formed by four $SU(2)_{2^{d+1}-2}$ anyons of topological charge $2^d-1$ instead, or any other variant of level and topological charges, for which a quqit still exists. 

\section{Conclusion}

Defining space-filling curves in dimension greater than $2$ is an interesting problem in its own. In this work, they are used as a framework to originate a representation of an integral point of a hypercube $2^n\times\dots\times 2^n$ of any dimension $d$ as an $n$-qutit, where $t=2^d$, on which novel representations for color multi-images are based. 

In a way, these novel representations introduce some intrinsic disorder. The brain cannot easily associate a correspondence between a given image data and the representation. 

Contrary to recent works of other authors which use representations with qutrits, the standard and usual data for the images (like traditional images'size or traditional number of bit planes) can be kept intact during our upgrade to qutits, $t\geq 3$, without needing to add any extra blank data. At the same time, our own model can easily be generalized to get also used with less traditional data formatting, for instance with images of any size $k^n\times k^n$ with $k\geq 3$. 

The original motivation for introducing these novel quantum representations for color multi-images using qutits was to decrease the storage space that is necessary to have on a quantum computer for the encryption and decryption of the images. This goal was clearly achieved, like was shown all throughout the paper. 

By generalizing the quantum baker map QBM so that it operates on a square of size $t^n\times t^n$ by shuffling bijectively two $n$-qutits, we have allowed for a much larger parameter space for quantum implementation of the QBM, thus avoiding brute-force attacks during the scrambling stage, or avoiding enlarging the number of secret keys arising from iteration parameters of the QBM. 
Some of our scrambling schemes use a novel concept which we name ``mixed scrambling". In some previous schemes by others and by the author, the quantum baker map gets applied on two $n$-qubits with parameters depending on  other data types not represented within the two $n$-qubits. For instance, bit plane independent scrambling of the pixel positions can be achieved with these controlled baker quantum gates, where the parameters of the QBM depend on the bit plane. In the present paper, due to the large parameter space for our generalized QBM, joint to mixed scrambling, quantum baker maps do not need to be applied independently in order to improve the security.  Again, this avoids having to transmit many secret keys to the receiver. 

In a previous scheme by the author, the images and the bit planes get scrambled in an identical way for each block of images. Some of the schemes developed here allow the blocks to be an integrant part of the scrambling process. 

By introducing the new concepts of ``mixed scrambling" and of ``mixed diffusion", we have permitted a broad number of schemes which can be designed according to the users' needs (type of data to encrypt and constraints of implementation). 

This work focused on the baker map for the scrambling of the qutits. However, other transforms having a quantum version using qubits could also be investigated in order to make them operational on qutits instead of qubits. 

Increasing the computational power of quantum computer using qutits, $t\geq 3$, instead of qubits might be the source of many more challenges in the years to come. Like for any other computational problem involving quantum computing, some approaches and hardware will be better suited than others for this specific field of quantum image encryption. 

$$\begin{array}{l}\end{array}$$

\textbf{Acknowledgements.} The author is pleased to heartfully thank Salavador E. Venegas-Andraca for introducing her to the field of quantum image encryption. She is quite happy to thank the mathematics department of the University of Southern California for very kind hospitality. 

$$\begin{array}{l}\end{array}$$

\textsc{Mathematics Department, University of Southern California, 3620 S. Vermont Avenue, Los Angeles, CA 90089, USA}\\

\textit{Email address:} \textit{clairelevaillant}$@$\textit{yahoo.fr}
\newpage
\begin{Large}Appendix.\end{Large} 3D projection of the sine chaotification of the 7D hyperchaotic system with respect to some of the coordinates.\\

On the plots below, the parameters of the sine chaotification are set to $(2; 3/2: 1; 7; 1/2; 6; 1)$. 
$$\begin{array}{l}\text{Last three coordinates:}\end{array}$$
\epsfig{file=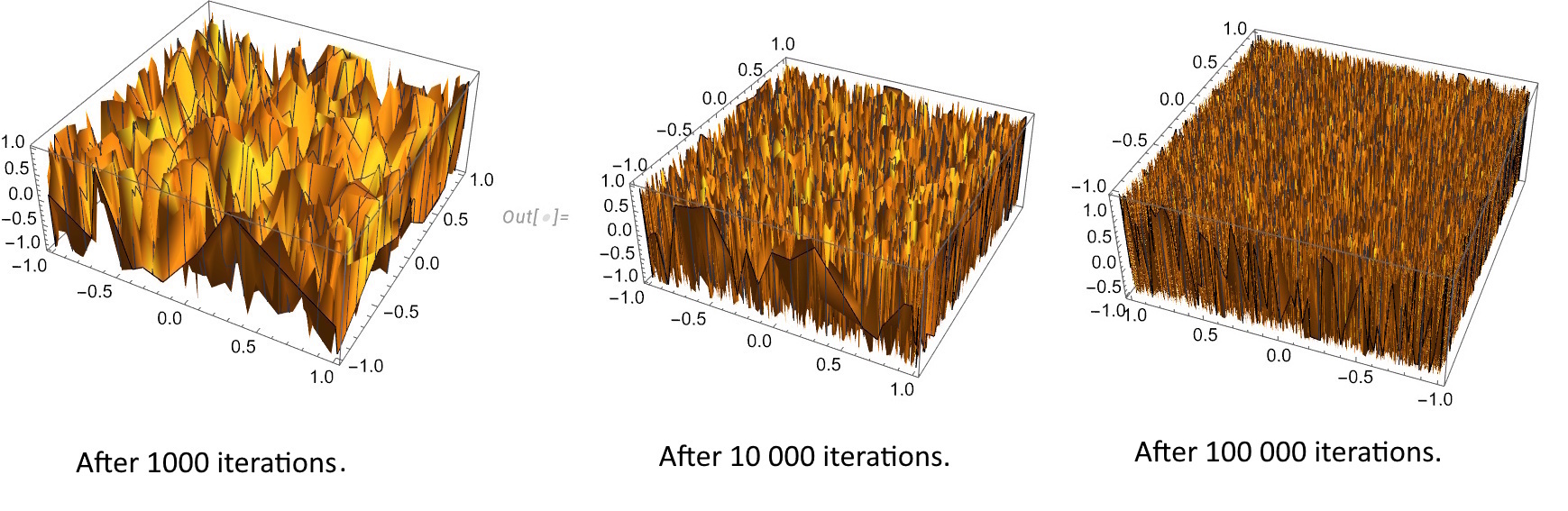, height=4cm}
$$\begin{array}{l}\text{First three coordinates:}\end{array}$$
\epsfig{file=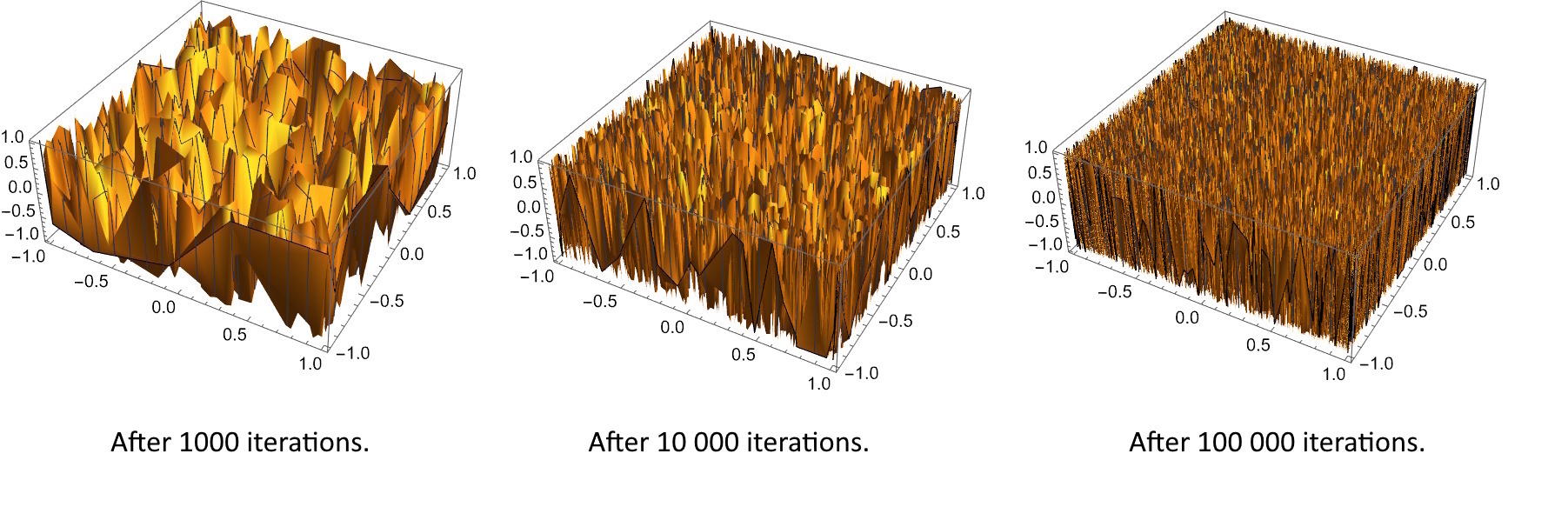, height=4cm}
$$\begin{array}{l}\text{Middle three coordinates:}\end{array}$$
\epsfig{file=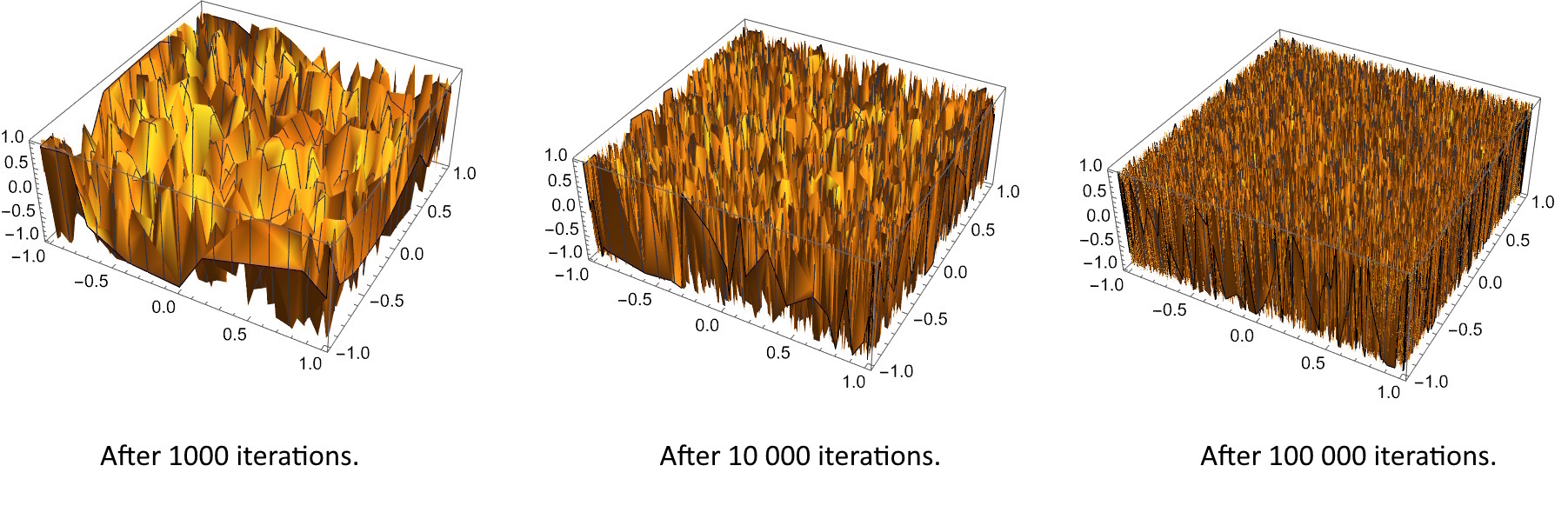, height=4cm}

\end{document}